\def\CleanRevisionMode{1}
\documentclass[preprint,12pt]{elsarticle}

\usepackage[sicmds,freestanding]{hepunits}
\usepackage{booktabs}
\usepackage{tabularx}
\usepackage{amssymb}
\DeclareUnicodeCharacter{03BD}{\ensuremath{\nu}}
\usepackage{subcaption}
\usepackage{float}
\usepackage{placeins}
\usepackage[mathlines]{lineno}
\usepackage{tikz}
\usetikzlibrary{arrows.meta,positioning,shapes.geometric,fit,decorations.markings,calc}
\usepackage[hidelinks]{hyperref}
\hypersetup{pdfsubject={paper\string_learned\string_fir}}

\usepackage{xcolor}
\colorlet{scientificblue}{blue}
\ifdefined\CleanRevisionMode\colorlet{blue}{black}\fi
\DeclareRobustCommand{\auditrevision}[1]{\textcolor{blue}{#1}}
\DeclareRobustCommand{\mauditrevision}[1]{\textcolor{blue}{#1}}
\newif\ifshowadaptiverevision
\showadaptiverevisiontrue
\DeclareRobustCommand{\adaptrevision}[1]{\ifshowadaptiverevision\textcolor{blue}{#1}\else#1\fi}
\makeatletter
\let\audit@originalseccntformat\@seccntformat
\def\@seccntformat#1{%
  \ifnum\pdfstrcmp{\detokenize{#1}}{subsection}=0
    \ifnum\value{section}=2
      \ifnum\value{subsection}<4
        \auditrevision{\csname the#1\endcsname.}\hskip 0.5em
      \else\audit@originalseccntformat{#1}\fi
    \else\ifnum\value{section}=4
      \ifnum\value{subsection}>2
        \auditrevision{\csname the#1\endcsname.}\hskip 0.5em
      \else\audit@originalseccntformat{#1}\fi
    \else\audit@originalseccntformat{#1}\fi\fi
  \else\audit@originalseccntformat{#1}\fi}
\makeatother
\makeatletter
\let\audit@originalbibitem\bibitem
\def\audit@bibitem{\@ifnextchar[\audit@bibitem@opt\audit@bibitem@noopt}
\def\audit@bibitem@prepare#1{%
  \color{black}%
  \def\audit@key{#1}%
  \def\audit@charrier{Charrier2019-ti}%
  \def\audit@kA{Correa2024-ap}%
  \def\audit@kB{NUTRIG2024}%
  \def\audit@kC{ARANext2024}%
  \def\audit@kD{RNOGCNN2025}%
  \def\audit@kE{Dorosti:2025ugq}%
  \def\audit@kF{Aksoy:2026hybrid}%
  \def\audit@kG{Aksoy:2026quantization}%
  \def\audit@kH{Ravn:2025likelihood}%
  \def\audit@kI{deBoone2026BOLT}%
  \def\audit@kJ{Fuchs2012AERA}%
  \ifx\audit@key\audit@charrier\color{blue}\else
    \ifx\audit@key\audit@kA\color{blue}\else
    \ifx\audit@key\audit@kB\color{blue}\else
    \ifx\audit@key\audit@kC\color{blue}\else
    \ifx\audit@key\audit@kD\color{blue}\else
    \ifx\audit@key\audit@kE\color{blue}\else
    \ifx\audit@key\audit@kF\color{blue}\else
    \ifx\audit@key\audit@kG\color{blue}\else
    \ifx\audit@key\audit@kH\color{blue}\else
    \ifx\audit@key\audit@kI\color{blue}\else
    \ifx\audit@key\audit@kJ\color{blue}\fi\fi\fi\fi\fi\fi\fi\fi\fi\fi
  \fi}
\def\audit@bibitem@after#1{%
  \def\audit@key{#1}%
  \def\audit@newkeyA{Charrier2019-ti}%
  \ifx\audit@key\audit@newkeyA\color{blue}\else\color{black}\fi}
\def\audit@bibitem@opt[#1]#2{%
  \audit@bibitem@prepare{#2}\audit@originalbibitem[#1]{#2}\audit@bibitem@after{#2}}
\def\audit@bibitem@noopt#1{%
  \audit@bibitem@prepare{#1}\audit@originalbibitem{#1}\audit@bibitem@after{#1}}
\let\bibitem\audit@bibitem
\makeatother

\journal{Astroparticle Physics}

\begin{document}
\begin{frontmatter}

\title{A dual-polarization whitened-template trigger for real-time radio detection of extensive air showers}

\author[siegen]{Qader Dorosti\corref{cor1}}
\ead{dorosti@hep.physik.uni-siegen.de}
\cortext[cor1]{Corresponding author}
\author[juelich]{Alperen Aksoy}
\author[juelich]{Ilja Bekman}
\author[siegen]{Markus Cristinziani}
\author[siegen]{Eric-Teunis de Boone}
\author[juelich,siegen]{Vesselin Dimitrov}
\author[juelich]{Chimezie Eguzo}
\author[ellab]{Stefan Heidbrink}
\author[juelich,due]{Stefan van Waasen}
\author[juelich]{Andre Zambanini}

\address[juelich]{Peter Gr\"unberg Institute -- Integrated Computing Architectures (ICA, PGI-4), Forschungszentrum J\"ulich GmbH, Germany}
\address[siegen]{Center for Particle Physics Siegen, Department Physik, Universit\"at Siegen, Germany}
\address[ellab]{Elektronikentwicklungslabor, Department Physik, Universit\"at Siegen, Germany}
\address[due]{Faculty of Engineering, Communication Systems, University of Duisburg-Essen, Germany}

\begin{abstract}
{\footnotesize \adaptrevision{Autonomous radio detection of extensive air showers requires reliable discrimination of their characteristic signals from transient interference, with decisions made in real time under stringent hardware constraints. We present a compact dual-polarization trigger developed using detector-folded air-shower simulations and measured radio backgrounds from the Auger Engineering Radio Array. The method combines the characteristic shape of air-shower pulses with measured background statistics to construct a filter that distinguishes air-shower signals from radio interference. On an independent validation dataset, it achieves \SI{96}{\percent} signal efficiency at a frame-equivalent background trigger rate of \SI{3.12}{\kHz}, substantially outperforming conventional amplitude thresholding and unwhitened template matching. Its robustness is further investigated using recordings acquired at different times from geographically distributed stations exposed to heterogeneous radio interference. Transferring the unchanged filter between these environments requires adjusting the trigger threshold to the recently measured background conditions to control the background trigger rate. With threshold updates every \SI{1}{\milli\second}, the achieved rates range from \SIrange[range-phrase={ to }]{35.2}{41.4}{\kHz} for a nominal target of \SI{35.7}{\kHz}, demonstrating effective rate control across the tested environments while maintaining sensitivity to air-shower signals. The compact trigger architecture meets the timing constraints of a field-programmable gate array at \SI{200}{\MHz}, supporting its feasibility for autonomous radio detection.}}

\end{abstract}

\begin{keyword}
radio detection \sep extensive air showers \sep real-time trigger \sep matched filter \sep field-programmable gate array \sep measured background
\end{keyword}
\end{frontmatter}

\section{Introduction}
\label{sec:intro}

Radio detection has become a mature technique for measuring extensive air showers induced by high-energy cosmic rays and neutrinos.  Modern digital radio arrays exploit the coherent radio emission produced by the geomagnetic and charge-excess mechanisms and can reconstruct air-shower observables such as arrival direction, energy, and shower maximum with high duty cycle~\cite{HUEGE20161,SCHRODER20171,geomEffect,AskEffect,Abdul_Halim2024-ll,Aab2016-fq}.  A reliable radio self-trigger is therefore attractive for large sparse arrays and for inclined showers, where the particle footprint can be strongly attenuated while the radio footprint remains large.

The trigger problem is difficult because the signal is short and the radio background is not purely thermal.  Field stations observe transient anthropogenic interference, narrow-band radio-frequency interference (RFI), and site-dependent changes in the noise environment.  Early radio self-trigger studies therefore combined thresholding, filtering, and data-acquisition constraints to keep false triggers manageable~ \auditrevision{\cite{Schmidt2011-xy,Torres_Machado2013-my,KELLEY2013133,Charrier2019-ti}}.  The practical requirement is not only classification accuracy.  A useful first-level trigger must operate at high trial rate, preserve sensitivity to weak pulses, and fit in the field-programmable gate array (FPGA) resources and latency available in autonomous stations.

Several nearby approaches have been explored.  Classical radio triggers use amplitude, envelope, band-pass, coincidence, or template-like quantities because they map naturally to firmware.  The NUTRIG first-level-trigger programme for the \auditrevision{Giant Radio Array for Neutrino Detection (GRAND)} studies template fitting and convolutional neural network (CNN)-based detection-unit triggering~\auditrevision{\cite{Correa2024-ap,NUTRIG2024}}.  ARA-Next proposes a radio-frequency system-on-chip (RFSoC)-based trigger architecture with more sophisticated trigger menus, including template-based and multi-trigger concepts~\auditrevision{\cite{ARANext2024}}.  \auditrevision{The Radio Neutrino Observatory in Greenland (RNO-G)} has investigated CNN filtering for radio-neutrino candidate selection~\auditrevision{\cite{RNOGCNN2025}}.  In the air-shower context, a fully convolutional radio self-trigger has demonstrated strong background rejection together with FPGA feasibility~\auditrevision{\cite{Dorosti:2025ugq}}, while a complementary hybrid approach combines neural waveform denoising and classification to retain near- and sub-threshold pulses under tight hardware constraints~\auditrevision{\cite{Aksoy:2026hybrid}}.  Studies of neural-network quantisation for scientific edge-computing applications further quantify the accuracy--resource trade-offs relevant to embedded inference~\auditrevision{\cite{Aksoy:2026quantization}}.

\auditrevision{Beyond these trigger developments, radio-signal analysis has also demonstrated the importance of accounting for correlations in measured noise. Band limitation, narrow-band interference, and the antenna and electronics response make neighbouring analog-to-digital converter (ADC) samples statistically dependent. These correlations can be characterized by a covariance matrix, whose diagonal elements describe the noise variance and whose off-diagonal elements describe correlations between samples at different time separations. Previous work has shown how incorporating noise correlations into likelihood-based analyses can improve radio-signal reconstruction and help distinguish cosmic-ray pulses from background noise~\cite{Ravn:2025likelihood}. This motivates investigating whether the same information can be exploited through a simpler filtering approach suitable for real-time triggering under the hardware constraints of autonomous stations.}

\adaptrevision{In this work, we develop and evaluate a compact radio trigger that combines the characteristic shape of air-shower pulses with measured background correlations. We construct and evaluate the trigger using detector-folded air-shower simulations and measured radio backgrounds from the Auger Engineering Radio Array (AERA)~\cite{Abdul_Halim2024-ll}, the radio-detector component of the Pierre Auger Observatory in Argentina~\cite{auger}. AERA comprises 153 autonomous stations operating in the \SIrange{30}{80}{\MHz} band.} The template response is evaluated on the two orthogonal high-gain polarization channels provided by the two crossed butterfly antennas of an AERA station, oriented north--south and east--west. In its use of expected air-shower pulse morphology at the detection-unit level, the method is related to the template-fitting trigger studied by GRAND/NUTRIG~\auditrevision{\cite{Correa2024-ap,NUTRIG2024}}. \auditrevision{Unlike that approach, one}  template incorporates both the expected pulse morphology and the measured AERA background covariance. The \auditrevision{filter operates} continuously, without \auditrevision{requiring} an amplitude-based pre-trigger \auditrevision{, and} the responses of both polarizations contribute to the trigger score. \adaptrevision{The resulting FPGA implementation processes one new dual-polarization sample pair per} clock cycle after the pipeline fills, \auditrevision{corresponding to an initiation interval of one (II=1).} 

\newpage
  \auditrevision{To investigate the performance of the proposed trigger beyond the original AERA benchmark, we use independent recordings from the Broadband Observatory of Lightning and Terrestrial Gamma-ray Flashes (BOLT), which retains the AERA detector hardware while enabling extended waveform acquisition~\cite{Fuchs2012AERA,deBoone2026BOLT}. These recordings allow us to examine the response of the unchanged whitened template to radio backgrounds measured at different times and locations. We first assess its performance on later data from the reference station without recalibrating the operating threshold. We then extend the study to other stations, where differences in local radio interference can substantially alter the background trigger rate. To accommodate these variations without modifying the learned filter, we investigate an operating threshold that adjusts to recent background measurements. The longer recordings also allow us to assess the temporal distribution of trigger decisions and their implications for downstream processing, which cannot be established from independent waveform segments alone.} 

Figure~\ref{fig:method_block} summarises the stage-1 trigger. \adaptrevision{The trigger is intended to select signal windows for subsequent station-level processing, multi-channel coincidence, or array-level event building. The following sections describe the datasets and trigger method, evaluate its performance across different stations, present the FPGA implementation, and discuss the physics interpretation of the results.}

\begin{figure}[b]
\centering
\resizebox{0.98\textwidth}{!}{%
\begin{tikzpicture}[
  node distance=8mm and 10mm,
  data/.style={draw=scientificblue!65!black, fill=scientificblue!8, align=center, minimum height=8mm, minimum width=22mm},
  process/.style={draw=teal!70!black, fill=teal!10, rounded corners, align=center, minimum height=8mm, minimum width=22mm},
  threshold/.style={draw=red!70!black, fill=red!9, rounded corners, align=center, minimum height=8mm, minimum width=22mm},
  arr/.style={-{Latex[length=2mm]}, thick}
]
\node[data] (adc) {N--S and E--W\\ADC streams};
\node[process, right=of adc] (fir0) {N--S polarization\\template scan};
\node[process, below=of fir0] (fir1) {E--W polarization\\template scan};
\node[process, right=13mm of fir0, yshift=-8mm] (pair) {local polarization\\combination};
\node[threshold, right=of pair] (thr) {score\\threshold};
\node[data, right=of thr] (trig) {trigger bit};
\draw[arr] (adc) -- (fir0);
\draw[arr] (adc) |- (fir1);
\draw[arr] (fir0) -- (pair);
\draw[arr] (fir1) -- (pair);
\draw[arr] (pair) -- (thr);
\draw[arr] (thr) -- (trig);
\end{tikzpicture}
}
\caption{Stage-1 trigger structure. The north--south and east--west high-gain ADC streams from the two crossed butterfly antennas are scanned with the same whitened pulse template. A local dual-polarization combination accepts pulse-like responses that appear in both polarizations within a small timing radius, and \adaptrevision{a score threshold} produces the trigger bit.}
\label{fig:method_block}
\end{figure}

\section{Datasets}
\label{sec:data}

\subsection{ \auditrevision{Reference benchmark and signal} construction}
 \label{sec:reference_benchmark}
 \label{sec:trace_construction}

The trigger \auditrevision{was constructed and calibrated using a reference benchmark combining measured radio background from AERA station 71 with simulated air-shower pulses propagated through the detector response. The measured background consists of synchronized traces from the two high-gain polarization channels of the station's crossed butterfly antennas, oriented north--south and east--west. The channels are digitized at a sampling rate of \SI{180}{\MHz}. Throughout this paper, we refer to these synchronized two-channel recordings as paired traces.} \auditrevision{The pure-signal library} is drawn from \auditrevision{the event-by-event AERA energy-scale simulation campaign~\cite{Huege2025-dm}, which} is based on  \num{844} high-quality \auditrevision{measured AERA cosmic-ray} events with reconstructed surface-detector energies \auditrevision{above} \SI{3e17}{\electronvolt}. For each  event, one proton- and one iron-induced  shower were simulated using \auditrevision{its reconstructed} geometry and calorimetric energy\auditrevision{, with} Sibyll~2.3d\auditrevision{, Ultrarelativistic Quantum Molecular Dynamics (UrQMD),} and  event-specific atmospheric profiles derived from \auditrevision{Global Data Assimilation System (GDAS)}. \auditrevision{The showers were propagated through the standard AERA detector and data-acquisition (DAQ) response using the calibration chain applied in that energy-scale study, producing paired detector-level station traces. The resulting pulse library retains the simulated detector-level amplitudes, including weak station responses that would not necessarily have been selected as radio detections in the measured AERA event sample.
}

Throughout this paper, \auditrevision{\emph{background} denotes a} measured noise-only \auditrevision{pair of polarization traces, \emph{pure signal} the} detector-folded \auditrevision{extensive air shower (EAS) pulse pair} before background addition, and \auditrevision{\emph{signal} the pulse pair injected into} measured background. \auditrevision{The paired channels retain their measured correlations} and  the relative timing \auditrevision{and amplitudes} predicted by the detector simulation. \auditrevision{A pulse pair is retained in the constructed sample if the peak in its weaker-polarization channel is at least one ADC count.}

\auditrevision{To form each signal event, a 256-sample pulse template drawn from the pure-signal library is added to a synchronized 2048-sample pair of measured-background traces. Its start position is selected uniformly from samples 64 through 1728, keeping the pulse at least 64 samples from either edge. The first and last 16 samples are then cropped from each trace, leaving 2016-sample analysis frames. Each resulting pair is one benchmark event: signal events contain the injected pulse pair in measured background, whereas background events contain measured background alone. For independent-frame background acceptance probability \(p_{\mathrm{frame}}\), the frame-equivalent background trigger rate is
} \begingroup
\makeatletter
\let\audit@oldtagform\tagform@
\def\tagform@#1{\audit@oldtagform{\auditrevision{#1}}}
\makeatother
\begin{equation}
\label{eq:ftr}
 f_{\mathrm{frame}}=p_{\mathrm{frame}}\frac{f_s}{2016}.
\end{equation}
\endgroup
This \auditrevision{conversion does not represent overlapping-window correlations or trigger timing.
}

\begin{figure}[tbp]
\centering
\includegraphics[width=0.82\textwidth]{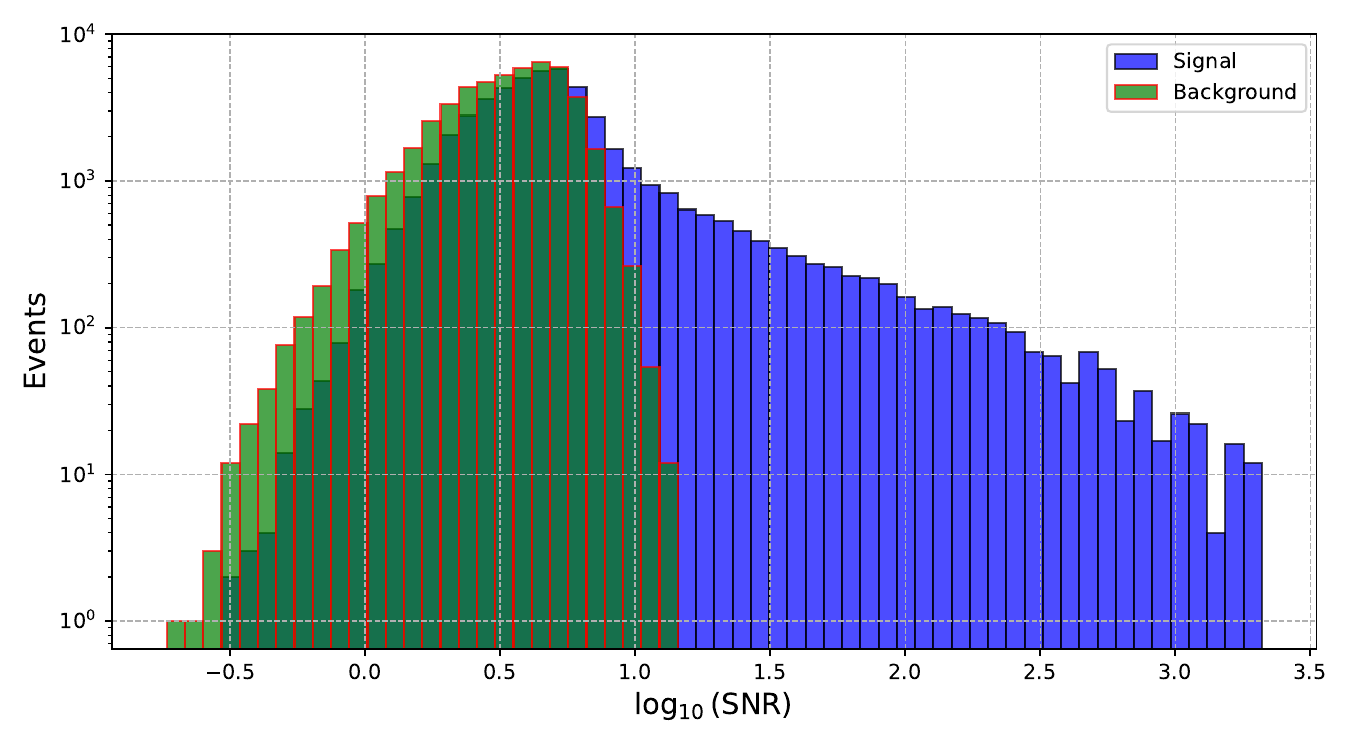}
\caption{Distribution of \(\log_{10}(\mathrm{SNR})\) for \auditrevision{signal} and \auditrevision{background} in \auditrevision{the reference benchmark, using} Equation~\eqref{eq:pair_snr_proxy}.}
\label{fig:snr_distribution}
\end{figure}

  \auditrevision{Figure~\ref{fig:snr_distribution} illustrates the range of signal strengths obtained by combining simulated air-shower pulses with measured background from the reference station.} The analysis-side \auditrevision{signal-to-noise ratio (SNR) proxy is
}\begingroup
\makeatletter
\let\audit@oldtagform\tagform@
\def\tagform@#1{\audit@oldtagform{\auditrevision{#1}}}
\makeatother
\begin{equation}
\label{eq:pair_snr_proxy}
\mathrm{SNR}=\frac{1}{\sigma_{\mathrm{pair}}^2}
\left(\frac{1}{|\mathcal{P}|}\sum_{i\in\mathcal{P}}\max_c|\mathcal{H}(x_c)_i|^2\right),
\end{equation}
\endgroup 
where \(\mathcal{H}\) \auditrevision{is the Hilbert transform, \(\mathcal{P}\) is the 65-sample pulse region centred on the larger pure-signal envelope maximum,} and \(\sigma_{\mathrm{pair}}^2\) is the mean \auditrevision{variance of the paired} measured-background \auditrevision{channels. For signal events,} \(x_c\) is the noisy signal trace. \auditrevision{For background events, the background trace is evaluated over the same pulse-centred support. An \(\mathrm{SNR}\) of unity then corresponds to a pulse-envelope power comparable to the mean background variance on that support.
}

  \auditrevision{The reference benchmark contains two classes, signal and background, divided into three disjoint splits:} \auditrevision{training contains 7000 events per class, validation set 1 contains 2000} events \auditrevision{per class, and validation set 2 contains} 2000  events per class \auditrevision{(Table~\ref{tab:dataset_definition}). Training constructs the pulse template and background covariance; validation set 1 supports architecture search} and threshold calibration; \auditrevision{validation set 2 is the original independent performance comparison.
The 11,000 measured-background frames were randomly assigned without replacement to the three splits. Frames were not screened according to how quiet or trigger-like they appeared}.

\begin{table}[tbp]
\centering
\small
\caption{ \auditrevision{Disjoint splits of the reference benchmark}. Each event contains \auditrevision{paired 2016-sample} north--south and east--west  traces .}
\label{tab:dataset_definition}
  \begin{tabular}{>{\raggedright\arraybackslash}p{0.22\textwidth}r>{\raggedright\arraybackslash}p{0.52\textwidth}}
 \toprule
Split & Events/class & Role  \\
\midrule
Training & 7000 & \auditrevision{Pulse-template and background-covariance construction} \\
Validation set 1 & 2000 & Architecture search and threshold calibration  \\
Validation set 2 & 2000 & Independent \auditrevision{reference performance evaluation} \\
\bottomrule
\end{tabular}
\end{table}

  \subsection{\auditrevision{Independent BOLT evaluation data}}
\label{subsec:multistation_backgrounds}

\auditrevision{Transient radio interference can produce short pulses that resemble air-shower signals, and the background differs between station locations. AERA recordings span many years and include background-triggered events, but their short, separate traces do not preserve the sequence of decisions made by a continuously operating trigger. Studying that sequence requires longer waveforms; testing the trigger in different interference environments requires stations across a wider area. BOLT provides both by repurposing AERA radio detectors, extending their waveform readout, and placing stations beyond the original AERA area~\cite{Fuchs2012AERA,deBoone2026BOLT}. Because BOLT reuses AERA station hardware, the detector-folded simulated pulses provide realistic signal inputs for evaluation with the BOLT recordings. We use later paired recordings from station 71 and stations 008, 042, 043, 088, and 137 to examine their measured backgrounds and trigger response, with the long records also supporting continuous replay of candidate bursts and queue occupancy.
}

\auditrevision{The BOLT evaluation uses later background recordings from station 71 and other stations, with their acquisition periods and available waveform durations summarized in Table~\ref{tab:station_context}. Battery and communication problems during the austral winter limited the availability of native BOLT recordings. These recordings support independent validation and background characterization without retraining the finite-impulse-response (FIR) filter.
}

\begin{figure}[tbp]
\centering
\includegraphics[width=0.88\textwidth]{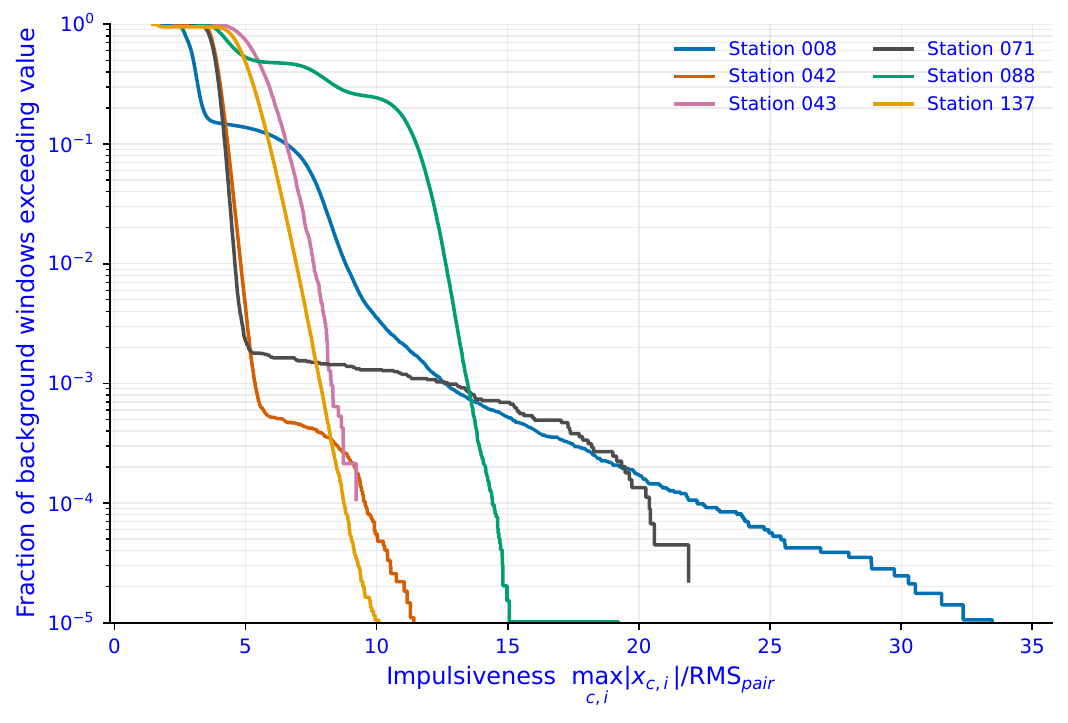}
\captionsetup{labelfont={color=blue}}
\caption{\auditrevision{Measured-background impulsiveness by station. Each curve gives the fraction of non-overlapping 2016-sample paired windows exceeding the value on the horizontal axis. The station 71 sample combines reference-benchmark background windows, later BOLT windows, and native BOLT recordings. Each curve is normalized by its own window count, so differences in available waveform duration affect how far into the tail it extends.}}
\captionsetup{labelfont={color=black}}
\label{fig:measured_background_impulsiveness}
\end{figure}

\begin{samepage}
\auditrevision{To compare large background excursions across stations, some of which can resemble air-shower pulses, we examine the paired recordings in non-overlapping 2016-sample windows, matching the analysis-frame length introduced above. For each window, we express the largest excursion relative to its background level through the impulsiveness measure
}\begingroup
\makeatletter
\let\audit@oldtagform\tagform@
\def\tagform@#1{\audit@oldtagform{\auditrevision{#1}}}
\makeatother
\begin{equation}
 \auditrevision{I=\frac{\max_{c,i}|x_{c,i}|}{\mathrm{RMS}_{\mathrm{pair}}},
}\end{equation}
\endgroup
\end{samepage}
\auditrevision{where \(x_{c,i}\) is the measured background sample in polarization channel \(c\) at sample \(i\) of the 2016-sample window. The paired RMS is calculated over both channels and the same window. Figure~\ref{fig:measured_background_impulsiveness} shows the fraction of windows at each station whose impulsiveness exceeds a given value. Station 088 has a broad tail of large excursions, station 043 shows consistently elevated values, and station 008 has occasional extreme excursions. Stations 042 shows fewer large relative excursions, whereas station 137 shows a broad elevated distribution approaching station 088: \SI{48}{\percent} of its background windows contain a peak at least five times that window's own background RMS (\SI{53}{\percent} at station 088, \SI{14}{\percent} at station 008, \SI{0.4}{\percent} at station 042), and the median peak-to-RMS ratio is 4.97, compared with 5.31, 3.01, and 3.89 at stations 088, 008, and 042. Because the available waveform durations differ, the curves extend to different minimum fractions.
}

\begin{table}[tbp]
\begingroup\color{blue}
\centering
\footnotesize
\setlength{\tabcolsep}{7pt}
\captionsetup{labelfont={color=blue}}
\caption{\auditrevision{Native BOLT recordings by station. Acquisition dates give the number of days and the span between the first and last record; waveform durations are totals over recorded traces, not continuous observing times.} \adaptrevision{The continuous evaluation uses the seventeen approximately one-second records identified in Section~\ref{subsec:continuous_bolt}.}}
\captionsetup{labelfont={color=black}}
\label{tab:station_context}
\begin{tabularx}{\textwidth}{lXrr}
\toprule
Station & Acquisition dates (2026) & Records & Total waveform duration \\
\midrule
008 & \SI{3}{days} (1--5 Oct) & 60 & \SI{3.18}{\second} \\
042 & \SI{7}{days} (3 Aug--5 Oct) & 53 & \SI{3.04}{\second} \\
043 & \SI{6}{days} (3 Aug--5 Sep) & 111 & \SI{0.105}{\second} \\
71  & \SI{12}{days} (3 Aug--6 Sep) & 268 & \SI{0.266}{\second} \\
088 & \SI{6}{days} (2 Sep--5 Oct) & 88 & \SI{2.20}{\second} \\
137 & \SI{3}{days} (5, 29–30 Oct) & 12 & \SI{11.70}{\second} \\
\bottomrule
\end{tabularx}
\endgroup
\end{table}

\auditrevision{For station 137, two recordings were excluded
from the continuous-trigger analysis because one of the two
polarization channels contained no valid data, while a third
recording was used for initial threshold calibration. The recordings sample distinct local settings across BOLT. Station 71, the reference site, lies in the medium-range cluster near pumping infrastructure, while station 088 is near power lines. Station 042 belongs to the core cluster and station 137 is remote; the different distributions at stations 042 and 043 provide another contrast among the measured backgrounds~\cite{deBoone2026BOLT}.
}


\section{Dual-polarization whitened-template trigger method}
\label{sec:method}

The trigger studied in this work is deliberately compact. It consists of one or more short FIR templates, independent streaming convolutions on the north--south and east--west polarization channels, a local polarization-combination rule, and a final threshold. The design is a whitened matched-filter-like trigger constrained from the outset for continuous FPGA implementation.

\subsection{Streaming trigger problem}
\label{sec:window_problem}

Let \(x_c[t]\) be the signed ADC value at sample \(t\) in channel \(c\in\{0,1\}\).  For a template of length \(L\), the per-channel response to template \(k\) is
\begin{equation}
\label{eq:channel_response}
  r_{c,k}(t) = \sum_{i=0}^{L-1} h_{k,i}x_c[t+i],
\end{equation}
where \(h_{k,i}\) are fixed template coefficients. The deployed implementation uses quantised integer coefficients, but the equations below are written in floating notation for clarity. The local magnitude response in a channel is
\begin{equation}
  a_c(t)=\max_k |r_{c,k}(t)|.
\end{equation}
The formulation allows a bank of \(K\) templates. The selected model uses \(K=1\) and \(L=16\), for which the channel response reduces to \(a_c(t)=|r_{c,1}(t)|\).

The trigger is evaluated for every valid sample position. A frame of \(N=2016\) samples gives \(N-L+1\) local trigger trials.  This convention matches a streaming FPGA use case: after the initial shift register is filled, the trigger can accept one new ADC pair per clock and compute one new local decision per clock if the synthesized design reaches initiation interval \(\mathrm{II}=1\).

\subsection{Whitened pulse-template construction}
\label{sec:fir_learning}

\begin{figure}[tbp]
\centering
\includegraphics[width=0.80\textwidth]{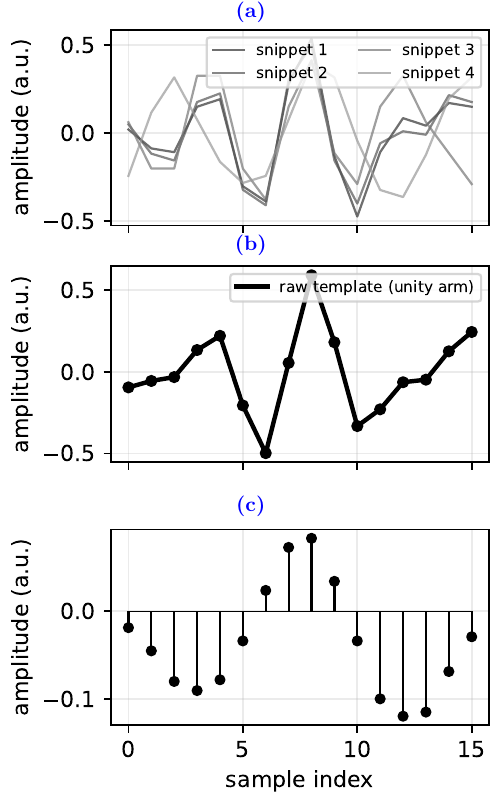}
\captionsetup{labelfont={color=blue}}
\caption{\auditrevision{Template construction: (a) aligned, mean-subtracted representative training pulse snippets; (b) their averaged pre-whitening pulse template; (c) the covariance-whitened 16-tap template used by the selected trigger.}}
\captionsetup{labelfont={color=black}}
\label{fig:template_construction}
\end{figure}

The template \auditrevision{construction turns a library of simulated detector-folded pulses into one short waveform that represents their common pulse shape, then weights that shape using correlations measured in the training background. For each of the 7000} pure-signal \auditrevision{events in the training split, the pulse maximum is located} independently in the \auditrevision{north--south and east--west channels. An event is omitted from template construction only when either maximum} is too close to the trace boundary \auditrevision{to extract the requested snippet}. Around each accepted channel \auditrevision{maximum}, an \(L\)-sample \auditrevision{waveform is extracted. Before averaging, we subtract each snippet's local mean to remove baseline shifts and align its polarity by making its largest-magnitude sample positive, so pulses of opposite sign do not cancel. We then normalize each snippet to unit length, preventing high-amplitude pulses from dominating the average. The resulting average is the single pulse-shape template used in the selected model (Figure~\ref{fig:template_construction}). The following equations give the exact operations. Each aligned snippet \(p_j\in\mathbb{R}^{L}\) is transformed to
} \begin{equation}
  u_j = \frac{p_j-\langle p_j\rangle}{\|p_j-\langle p_j\rangle\|_2},
\end{equation}
  \auditrevision{where \(\langle p_j\rangle\) is the snippet mean} and \(\|\cdot\|_2\) \auditrevision{is its} Euclidean norm. The first pulse template \(s_1\) is the \auditrevision{normalized average of the} aligned snippets. 
  \auditrevision{For the multi-template architectures evaluated in Section~\ref{sec:hardware_aware_model_selection}, additional pulse shapes are extracted using singular value decomposition (SVD) to capture the main variations among the training pulses.}

The \auditrevision{next step estimates the background correlations using the paired background traces in the 7000-event} training split. For each \auditrevision{channel trace, its mean is removed, then the average product of samples separated by lag \(d\) is computed. This gives the lag covariance
} \begin{equation}
  \gamma_d =
  \mathbb{E}_{q,t}\left[
    b_q[t]\,b_q[t+d]
  \right],
\end{equation}
where \(b_q[t]\) is the mean-subtracted ADC value of background trace/channel \(q\), and the \auditrevision{average runs over the training traces, both} channels, and valid sample positions. \auditrevision{The lag values form} an \(L\times L\) Toeplitz covariance matrix,
\begin{equation}
  (C_n)_{ij} = \gamma_{|i-j|} + \lambda\gamma_0\delta_{ij},
  \qquad i,j=0,\ldots,L-1 ,
\end{equation}
where \(\delta_{ij}\) is the Kronecker delta and \(\lambda=10^{-3}\) in the reference configuration. The \auditrevision{small diagonal term stabilizes the inverse against numerical sensitivity. Samples separated by the same number of time steps have the same estimated covariance, giving \(C_n\) its Toeplitz form}.

 \auditrevision{Whitening gives greater weight to pulse-shape components that are unusual in the measured background and less weight to components that commonly occur as correlated noise.} For each pulse-shape template \(s_k\), the \auditrevision{unnormalized} whitened template is obtained by solving
 \begin{equation}
\label{eq:whitened_template}
  \tilde h_k = C_n^{-1}s_k .
\end{equation}
It is then normalised to unit standard deviation under the measured background covariance,
\begin{equation}
  h_k =
  \frac{\tilde h_k}{\sqrt{\tilde h_k^T C_n \tilde h_k}} .
\end{equation}
\auditrevision{This normalization accounts for the background noise level and correlations, providing a consistent scale for calibrating the trigger threshold.} \auditrevision{The covariance is used offline to compute the fixed FIR coefficients \(h_k\). During operation, the FPGA applies those coefficients and evaluates the score and threshold.}

 \subsection{Dual-polarization score}
\label{sec:paired_score}

The two polarization channels are combined after the template response. This exploits the temporal correspondence of pulse-like morphology in the north--south and east--west components, while isolated background excursions in one polarization are not promoted by averaging. For a timing radius \(R\), define the local partner maxima
\begin{equation}
  m_0(t)=\max_{\Delta\in[-R,R]} a_0(t+\Delta),
  \qquad
  m_1(t)=\max_{\Delta\in[-R,R]} a_1(t+\Delta).
\end{equation}
Two polarization-combination rules were studied. The dual-polarization coincidence rule is
\begin{equation}
\label{eq:coincidence_score}
  S_{\mathrm{coinc}}(t)
  =
  \max\left[
    \min\{a_0(t),m_1(t)\},
    \min\{a_1(t),m_0(t)\}
  \right],
\end{equation}
which requires both polarizations to show a response within the radius. The polarization-sum rule is
\begin{equation}
\label{eq:sum_score}
  S_{\mathrm{sum}}(t)
  =
  \max\left[
    a_0(t)+m_1(t),
    a_1(t)+m_0(t)
  \right].
\end{equation}
The selected reference model uses Equation~\eqref{eq:sum_score} with \(R=8\) samples.  The score assigned to a full frame is
\begin{equation}
  S_{\mathrm{frame}} = \max_t S_{\mathrm{sum}}(t).
\end{equation}
The same frame score is used in software validation and in the high-level synthesis (HLS) implementation.

\subsection{Hardware-aware architecture selection}
\label{sec:hardware_aware_model_selection}

\auditrevision{Each tested architecture specifies a template length \(L\), a number of templates \(K\), and a rule for combining the two polarization responses. These three choices vary in the search; the coincidence radius and covariance regularization remain fixed (Table~\ref{tab:hpo_search_space}). The training split supplies the pulse templates and background covariance. We compare architectures on \emph{validation set 1} and leave \emph{validation set 2} untouched for the independent performance evaluation.}

\begin{table}[tbp]
\centering
\small
\captionsetup{labelfont={color=blue}}
\caption{\auditrevision{Architecture-selection settings: varied choices, fixed construction settings, and selection criteria.}}
\captionsetup{labelfont={color=black}}
\label{tab:hpo_search_space}
  \begin{tabularx}{\textwidth}{>{\raggedright\arraybackslash}p{0.25\textwidth}>{\raggedright\arraybackslash}p{0.20\textwidth}X}
 \toprule
Quantity & Values used & Role in the \auditrevision{comparison} \\
\midrule
\multicolumn{3}{l}{\auditrevision{\emph{Varied architecture choices}}} \\
Template length \(L\) & 16, 24, 32 samples & Controls the FIR window and the pulse-morphology information available to the trigger. \\
Number of templates \(K\) & 1, 2, 3, 4 & Controls the number of pulse-shape modes in the template bank. \\
Polarization-combination rule & polarization sum, dual-polarization coincidence & Tests whether efficiency is maximised by adding coincident polarization responses or by requiring a stricter coincidence. \\
\addlinespace[0.3em]
\multicolumn{3}{l}{\auditrevision{\emph{Fixed construction settings}}} \\
Coincidence radius \(R\) & 8 samples & Fixed timing tolerance between the two high-gain channels. \\
Covariance regularisation \(\lambda\) & \(10^{-3}\) & Fixed stabilisation term for the whitening covariance. \\
Template and covariance statistics & 7000 training events per class & Fixed sample size used to construct the templates and measured-background covariance. \\
\addlinespace[0.3em]
\multicolumn{3}{l}{\auditrevision{\emph{Selection criteria}}} \\
Reference \auditrevision{background-frame acceptance} & \(p_\star=10^{-2}\) & \auditrevision{Common background level for architecture comparison.} \\
Minimum \auditrevision{signal} efficiency & \auditrevision{\(\geq 0.90\)} & \auditrevision{Requirement before minimizing the hardware-cost proxy}. \\
Hardware-cost proxy & \(2KL\) multipliers & Estimated two-channel FIR multiplier count used for architecture ranking. \\
\bottomrule
  \end{tabularx}
 \end{table}

\auditrevision{An analysis window is accepted when its score exceeds the threshold. We call each accepted window a candidate and define the candidate rate as the number of candidates per second. Background acceptance and signal efficiency are the fractions of background and signal frames, respectively, that exceed the same trigger threshold. For each architecture, we choose a threshold corresponding to \SI{1}{\percent} background acceptance (20 of 2000 background frames) in \emph{validation set 1}. The estimated multiplier count is
}\begin{equation}
  C_{\mathrm{mult}} = 2 K L ,
\end{equation}
where the factor of two accounts for the two input channels. This \auditrevision{counts constant-coefficient FIR multipliers before synthesis-level sharing or pruning. Among architectures with at least \SI{90}{\percent} signal efficiency at the common \SI{1}{\percent} background acceptance, we select the lowest multiplier count, breaking ties by higher efficiency. If none reaches \SI{90}{\percent}, we select the highest-efficiency architecture. The \(K=1\), \(L=16\) polarization-sum rule reaches 0.95 efficiency at this comparison point. Its cost of 32 multipliers is the minimum in the search, and it has higher efficiency than the coincidence rule at the same cost (Table~\ref{tab:hpo_results}).}

\auditrevision{The \SI{1}{\percent} background acceptance is used only for architecture selection. The final operating threshold is calibrated separately in Section~\ref{subsec:validation_performance}.}

 \begin{table}[tbp]
\centering
\small
\captionsetup{labelfont={color=blue}}
\caption{\auditrevision{Single-template (\(K=1\)) architectures at \SI{1}{\percent} background-frame acceptance on validation set 1. The search also included \(K=2,3,4\); the table shows the minimum-cost comparisons.}}
\captionsetup{labelfont={color=black}}
\label{tab:hpo_results}
  \begin{tabularx}{\textwidth}{>{\raggedright\arraybackslash}Xccc}
 \toprule
\auditrevision{Polarization rule} & \auditrevision{Length (samples)} & Multipliers & \auditrevision{Signal efficiency} \\
\midrule
\auditrevision{Polarization sum} & 16 & 32 & \auditrevision{0.95} \\
\auditrevision{Dual-polarization coincidence} & 16 & 32 & \auditrevision{0.93} \\
\auditrevision{Polarization sum} & 24 & 48 & \auditrevision{0.99} \\
\auditrevision{Dual-polarization coincidence} & 24 & 48 & 0.98 \\
\auditrevision{Polarization sum} & 32 & 64 & 1.00 \\
\auditrevision{Dual-polarization coincidence} & 32 & 64 & 0.99 \\
\bottomrule
  \end{tabularx}
 \end{table}

\section{Trigger performance}
\label{sec:performance}

\auditrevision{We first compare the selected trigger with conventional
methods on the independent reference benchmark. We then test its
performance on later recordings from the reference station and
investigate continuous operation across different stations,
including threshold adaptation and downstream processing demands.}

\subsection{Reference benchmark performance}
\label{subsec:validation_performance}

\auditrevision{To compare the trigger methods at approximately the same background level, we choose \SI{4}{\percent} background-frame acceptance on
\emph{validation set 1} (80 of 2000 frames). This balances
background rejection against the statistical precision of
threshold calibration. The corresponding frame-equivalent candidate rate is \SI{3.57}{\kHz}, approximately \SI{5}{\percent} of the \SI{71.4}{\kHz} processing capacity of the hybrid neural denoiser and classifier~\cite{Aksoy:2026hybrid}, which we use as an example downstream processing stage. Each method's threshold is calibrated on \emph{validation set 1}
and applied unchanged to \emph{validation set 2}.}

\begin{figure}[tbp]
\centering
\includegraphics[width=0.82\textwidth]{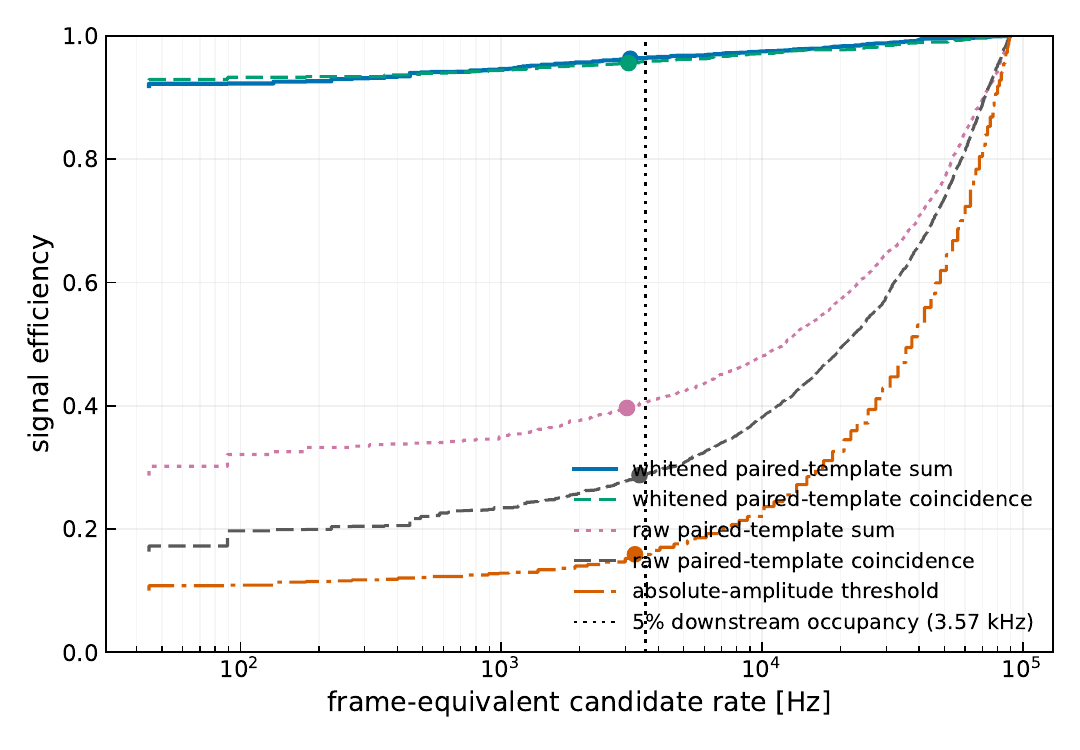}
\captionsetup{labelfont={color=blue}}
\caption{Independent \emph{validation set 2} efficiency versus frame-equivalent candidate rate. The circular markers indicate the calibrated operating points listed in Table~\ref{tab:operating_points}. The vertical line indicates the \SI{3.57}{\kHz} reference operating point. \auditrevision{At the same candidate rate, the whitened-template curves lie above the raw-template and amplitude baselines; the sum and coincidence curves are similar.}}
\captionsetup{labelfont={color=black}}
\label{fig:final_roc}
\end{figure}

Figure~\ref{fig:final_roc} shows the independent \emph{validation set 2} rate curves. The proposed trigger is compared with four controlled baselines: an absolute-amplitude threshold; the same pulse template without covariance whitening, using either polarization sum or coincidence; and the whitened template with the stricter coincidence rule. \auditrevision{These comparisons test pulse-template matching and covariance whitening, and compare the two polarization-combination rules.} The circular marker on each curve is the operating point obtained by calibrating that method's threshold on \emph{validation set 1} and applying it unchanged to \emph{validation set 2}; the corresponding values are listed in Table~\ref{tab:operating_points}.

\begin{table}[tbp]
\centering
\footnotesize
\captionsetup{labelfont={color=blue}}
\caption{\emph{Validation set 2} performance using thresholds calibrated on \emph{validation set 1}. Each \auditrevision{confidence interval (CI)} is a two-sided \SI{95}{\percent} Clopper--Pearson interval, and displayed values are rounded to two decimal places. Rate intervals propagate the binomial interval through Equation~\eqref{eq:ftr}.}
\captionsetup{labelfont={color=black}}
\label{tab:operating_points}
\begin{tabularx}{\textwidth}{>{\raggedright\arraybackslash}Xccc}
\toprule
Trigger & \shortstack{Efficiency\\(\SI{95}{\percent} CI)} & \auditrevision{\shortstack{Frame-equivalent\\candidate rate [\si{\kHz}]\\(\SI{95}{\percent} CI)}} & Threshold \\
\midrule
Whitened dual-polarization sum & \(\mauditrevision{0.96}\;(\mauditrevision{0.95},\mauditrevision{0.97})\) & \(\mauditrevision{3.12}\;(\mauditrevision{2.44},\mauditrevision{3.93})\) & \auditrevision{19.56} \\
Whitened dual-polarization coincidence & \(0.96\;(0.95,\mauditrevision{0.96})\) & \(\mauditrevision{3.08}\;(\mauditrevision{2.41},\mauditrevision{3.88})\) & \auditrevision{6.68} \\
Raw dual-polarization sum & \(\mauditrevision{0.40}\;(\mauditrevision{0.37},\mauditrevision{0.42})\) & \(\mauditrevision{3.04}\;(\mauditrevision{2.37},\mauditrevision{3.83})\) & \auditrevision{268.97} \\
Raw dual-polarization coincidence & \(\mauditrevision{0.29}\;(\mauditrevision{0.27},\mauditrevision{0.31})\) & \(\mauditrevision{3.39}\;(\mauditrevision{2.68},\mauditrevision{4.23})\) & \auditrevision{119.02} \\
Absolute-amplitude threshold & \(0.16\;(\mauditrevision{0.14},0.18)\) & \(3.26\;(2.56,4.08)\) & 266 \\
\bottomrule
\end{tabularx}
\end{table}

\auditrevision{At the selected trigger's fixed score threshold,} \emph{validation set 1} gives \(\mauditrevision{1932}/2000=0.97\) efficiency and \(80/2000\) accepted background frames, while \emph{validation set 2} gives \(\mauditrevision{1926}/2000=\mauditrevision{0.96}\) and \(\mauditrevision{70}/2000\), respectively. \auditrevision{The independent split shows similar efficiency and background acceptance at the unchanged threshold.} With 2000 background frames, one accepted frame corresponds to \SI{44.64}{\Hz}; the \SI{95}{\percent} rate intervals in Table~\ref{tab:operating_points} quantify this finite-sample uncertainty.

\FloatBarrier
\subsection{\auditrevision{Fixed-threshold signal efficiency on later station 71 backgrounds}}
 \label{subsec:later071_robustness}

\auditrevision{The independent validation above uses frames from the reference benchmark. Here we test whether the selected trigger still detects pulses in later backgrounds from the same station with its filter and threshold fixed.}

\begin{figure}[tbp]
\centering
  \includegraphics[width=0.82\textwidth]{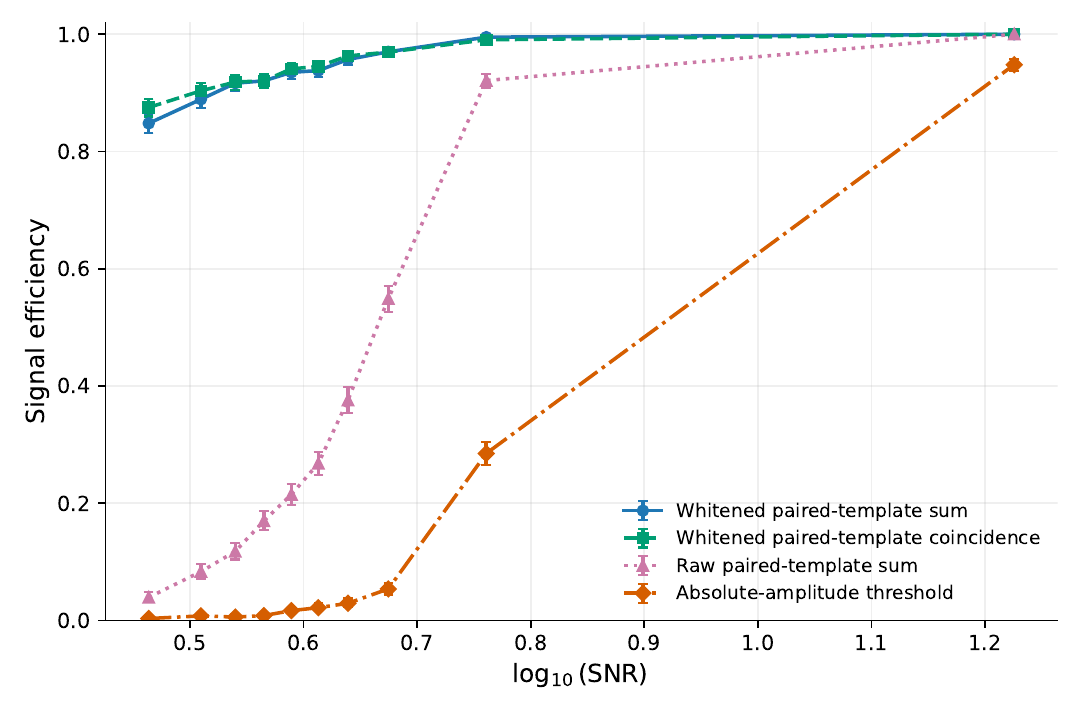}
 \captionsetup{labelfont={color=blue}}
\caption{\auditrevision{Fixed-threshold signal efficiency versus the analysis-side SNR proxy for 20,000 injections into later BOLT backgrounds from station 71. The detector-folded pulse library and validation-set-1 thresholds are unchanged from the reference analysis. Points use ten equal-population bins; vertical bars are exact two-sided} \SI{95}{\percent} Clopper--Pearson intervals and horizontal bars show bin extent.}
\captionsetup{labelfont={color=black}}
\label{fig:efficiency_snr}
\end{figure}

\auditrevision{We inject 20,000 unique held-out detector-folded pulses from the same library into non-overlapping windows drawn from 268 later BOLT records at station 71, acquired on separate days from early August to early September 2026. Figure~\ref{fig:efficiency_snr} compares the selected trigger with three baselines. Each uses its validation-set-1 score threshold without retraining or recalibration. We plot efficiency in ten equal-population bins of the SNR proxy defined in Equation~\eqref{eq:pair_snr_proxy}, with exact binomial \SI{95}{\percent} intervals. For these bins, the background variance is estimated from the later records.}

\auditrevision{In the lowest SNR bin, the selected whitened paired-template sum accepts 1696 of 2000 injected pulses (\SI{84.80}{\percent}; \SI{95}{\percent} CI \SIrange{83.15}{86.35}{\percent}). Efficiency rises to \SI{88.9}{\percent} in the next bin and approaches unity for stronger signals. Both whitened methods remain more efficient at low SNR than the raw-template and absolute-amplitude baselines. The later recordings thus provide an independent fixed-threshold check in which the whitened methods retain their advantage, while the weakest pulses remain the most difficult.}

\FloatBarrier
\subsection{\adaptrevision{Performance on continuous BOLT recordings}}
\label{subsec:continuous_bolt}

\adaptrevision{The preceding evaluations establish discrimination on independent waveform segments and sensitivity to pulses in later background from the reference station. We now examine continuous operation of the unchanged whitened template on seventeen approximately one-second recordings from stations 008, 042, 088 and 137. For station 137, two recordings were excluded because one
polarization channel contained no valid data, while a third
was reserved for initial threshold calibration.Here, detection performance must be considered together with the background trigger rate and the timing of accepted windows. A mean rate compatible with downstream processing can still contain short bursts that temporarily exceed its capacity.} \adaptrevision{We therefore examine several operating rates, including higher loads, to assess the trade-off between signal efficiency and downstream processing demands.}

\adaptrevision{The continuous recordings are analysed in 2048-sample windows.
Consecutive windows can overlap, allowing the same part of the waveform
to be examined more than once. Overlap can reduce the effects of window boundaries on pulse detection, but can also produce multiple trigger candidates from the same transient.
We evaluate overlaps of \SI{0}{\percent}, \SI{10}{\percent}, \SI{30}{\percent}, \SI{40}{\percent}, and \SI{50}{\percent} to examine
the trade-off between signal detection efficiency and downstream
processing demands.}

\adaptrevision{To accommodate variations in the measured radio background, the operating threshold is updated approximately every \SI{1}{\milli\second}. After every completed interval, the distribution of trigger scores measured during that interval determines the threshold for the next. The threshold is chosen so that applying it to the preceding measurements would accept no more windows than allowed by the target background rate. The first interval uses a threshold calibrated on earlier independent background recordings from the same station, including short recordings where a suitable preceding long recording was unavailable. Each recalculated threshold applies only to subsequent data, so no update uses future measurements. Because the background can change between intervals, the achieved rate can differ from the nominal target.}

\adaptrevision{Signal efficiency is evaluated with 2000 held-out detector-folded pulse pairs injected into measured background segments from each recording. Pulse positions are chosen uniformly relative to the analysis-window boundaries. The same injected signals and background segments are used for all rate targets and overlap settings to ensure a consistent comparison. A successful detection requires at least one accepted window whose maximum template score lies within 32 samples, approximately \SI{0.18}{\micro\second}, of the injected pulse-envelope maximum. The corresponding background waveform is checked for triggers within the same 32-sample interval. If it already produces a trigger within that interval, the event is counted as a missed detection. This prevents an interference trigger near the injection position from being mistaken for detection of the air-shower pulse.}

\begin{figure}[!t]
\centering
\includegraphics[width=\textwidth]{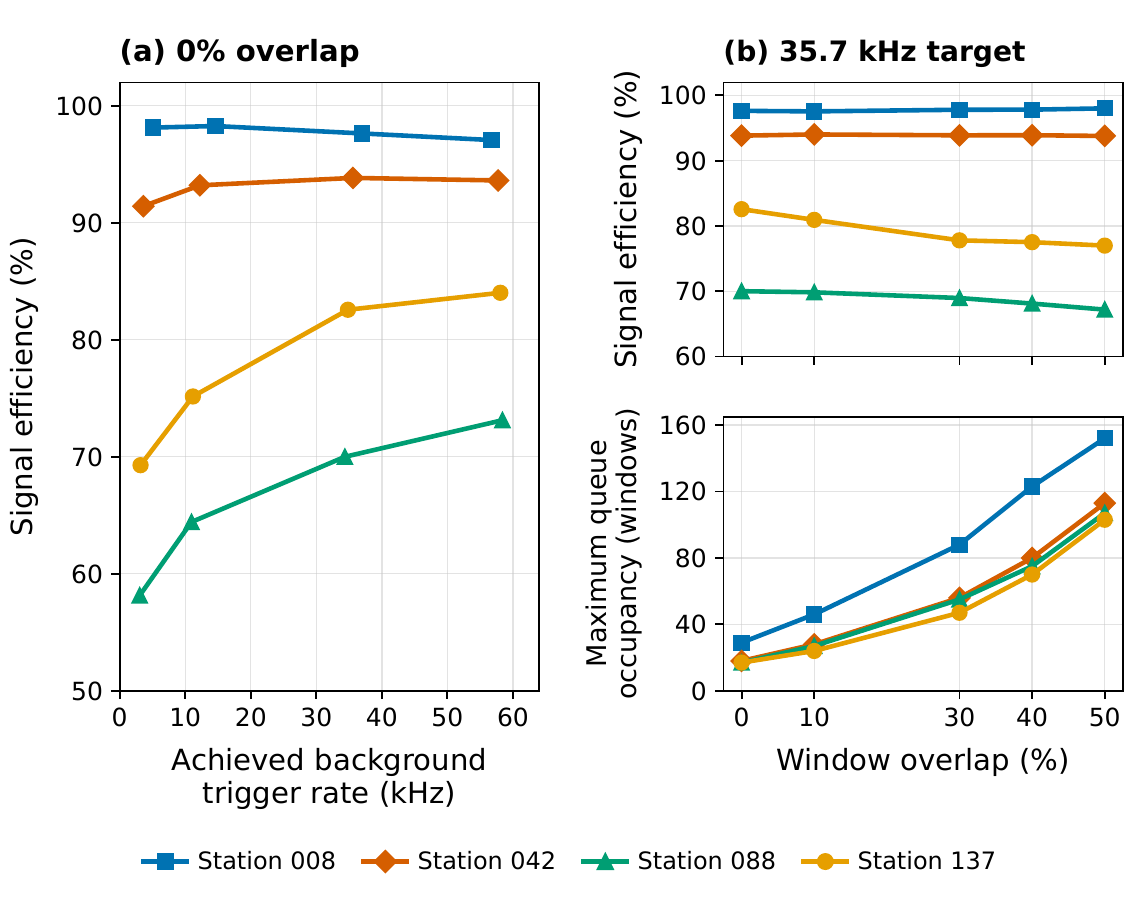}
\captionsetup{labelfont={color=blue}}
\caption{\adaptrevision{Performance of the whitened-template trigger with unchanged FIR coefficients and causal threshold updates at approximately \SI{1}{\milli\second} intervals on seventeen approximately one-second BOLT recordings, comprising three each from stations 008 and 042, two from station 088, and nine from station 137. (a) Pulse-attributed signal efficiency versus achieved background trigger rate at \SI{0}{\percent} overlap. Curves give averages over each station's recordings at nominal targets of \SIlist[list-final-separator={, and },list-units=single]{3.57;11.9;35.7;59.5}{\kHz}. (b) Signal efficiency (upper plot) and maximum downstream queue occupancy (lower plot) versus overlap at the nominal \SI{35.7}{\kHz} target. The upper curves give station averages, and the lower curves give the largest waiting count among recordings at each station. The queue occupancy excludes the window being processed,
and overlapping accepted windows count separately. Achieved rates vary with overlap, so the results compare a common nominal target rather than equal background rates.}}
\captionsetup{labelfont={color=black}}
\label{fig:multistation_overlap_stride}
\end{figure}

\adaptrevision{Figure~\ref{fig:multistation_overlap_stride}(a) shows the signal efficiency against the achieved background trigger rate at \SI{0}{\percent} window overlap. At the nominal \SI{35.7}{\kHz} target, corresponding to half the processing capacity of the example downstream stage introduced in Section~\ref{subsec:validation_performance},
individual recordings achieve \SIrange{34.1}{38.7}{\kHz}, illustrating why the measured rate is used on the horizontal axis. Station 008 maintains approximately \SIrange{97}{98}{\percent} efficiency across the tested range. Station 042 improves modestly before reaching approximately \SI{94}{\percent}. Stations 088 and 137 show stronger rate dependence, with efficiencies increasing from approximately \SI{58}{\percent} to \SI{73}{\percent} and from \SI{69}{\percent} to \SI{84}{\percent}, respectively. Adjusting the threshold therefore supports operation of the unchanged filter across these measured environments, but does not produce uniform detection efficiency between stations.}

\adaptrevision{The lower efficiency at station 088 occurs in recordings with substantial impulsive activity, as characterized in Section~\ref{subsec:multistation_backgrounds}. Station 042 has fewer large relative excursions, while station 008 retains high efficiency despite occasional extreme excursions. However, the impulsiveness measure alone does not capture the temporal structure or polarization characteristics of background interference, which can also influence the response of the whitened-template trigger and its detection efficiency. Station 137 further illustrates this contrast. Although its
impulsiveness distribution approaches that of station 088, its
mean efficiency at the \SI{35.7}{\kHz} target (\SI{83}{\percent}) falls
between those of stations 042 (\SI{94}{\percent}) and 088 (\SI{70}{\percent}).
Efficiencies across its individual recordings range from
\SI{66}{\percent} to \SI{91}{\percent}, highlighting substantial variation even
within the same station.}

\adaptrevision{The timing of accepted windows matters because the downstream stage described in Ref.~\cite{Aksoy:2026hybrid} processes one window approximately every \SI{14}{\micro\second}. Although the nominal \SI{35.7}{\kHz} target is half this service capacity, overlapping windows can contain the same transient and produce several closely spaced trigger decisions, each requiring downstream processing. The spacing between neighboring windows decreases from \SI{11.38}{\micro\second} without overlap to \SI{5.69}{\micro\second} at \SI{50}{\percent} overlap, allowing accepted windows to arrive faster than they can be processed. We use each accepted window's final-sample time as its arrival time and process windows in arrival order, allowing them to wait until the processor becomes available.}

\adaptrevision{Figure~\ref{fig:multistation_overlap_stride}(b) shows that signal efficiency changes little at stations 008 and 042 and decreases at station 088 as overlap increases, while waiting demand increases at all four stations. Station 137 behaves like station 088
in efficiency, decreasing from \SI{83}{\percent} to \SI{77}{\percent} between \SI{0}{\percent} and \SI{50}{\percent} overlap,
while its largest queue occupancy rises from 17 to 103 and stays below the
four-station maximum at every overlap. Across the tested recordings, the largest number waiting rises from 29 without overlap to 123 at \SI{40}{\percent} overlap and 152 at \SI{50}{\percent} overlap. Achieved background rates across these settings span \SIrange{34.1}{48.7}{\kHz} despite the common nominal target. These operating points therefore compare a common target rather than equal achieved rates and do not establish a universal optimum. Temporary bursts of accepted windows can require buffering even when the average background trigger rate remains below the downstream processing capacity. Together with the adaptive-threshold results, the queue analysis provides a basis for estimating the processing and buffering requirements of an autonomous radio trigger under measured background conditions.}

\FloatBarrier

\subsection{Example trigger decisions}
\label{subsec:example_trigger_decisions}

\begin{figure}[p]
\centering
  \includegraphics[width=0.80\textwidth]{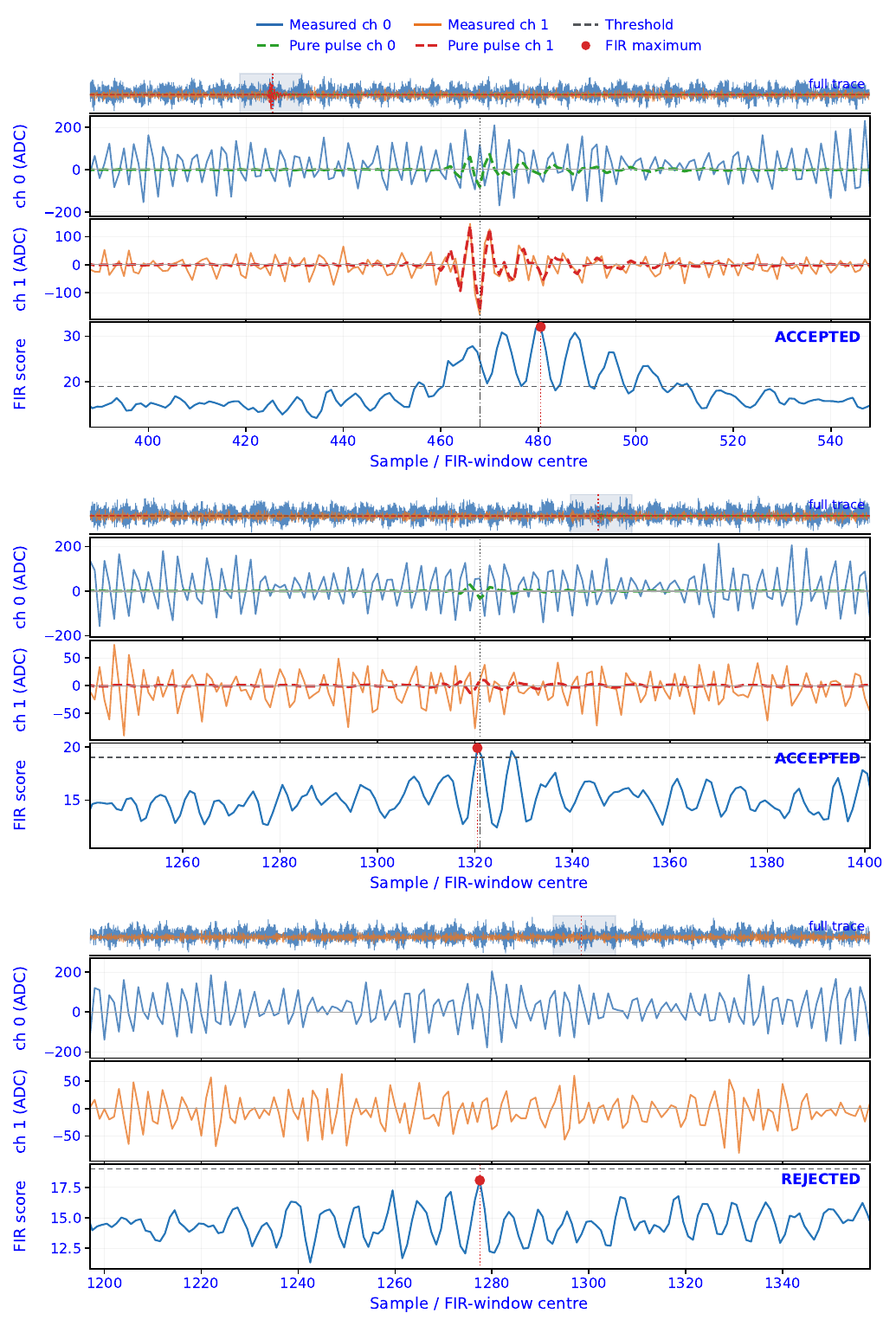}
 \captionsetup{labelfont={color=blue}}
\caption{Representative strong-signal, near-threshold signal, and rejected-background decisions. Upper panels show both polarization traces, with injected pulses overlaid for signal events; lower panels show the dual-polarization template score, \auditrevision{the calibrated trigger threshold} (dashed line), and frame maximum (red marker).}
\captionsetup{labelfont={color=black}}
\label{fig:example_trigger_decisions}
\end{figure}

\auditrevision{Figure~\ref{fig:example_trigger_decisions} illustrates the trigger response to a strong injected pulse, a weak injected pulse, and a background-only frame.} In each example, the upper panel contains the two measured polarization traces; for signal events, the detector-folded pure pulses are overlaid at their injection position. The lower panel shows the local dual-polarization template score, with the horizontal dashed line marking the fixed trigger threshold and the red marker identifying the frame maximum.

The strong signal produces a compact score peak with a clear margin above threshold at the injected-pulse position. The weak signal is also accepted, but its maximum lies only slightly above threshold, illustrating operation near the decision boundary. In contrast, the background-only example contains several sizeable ADC excursions but its maximum template score remains below threshold. 

\section{Firmware generation and \auditrevision{register-transfer-level (RTL)} validation}
\label{sec:firmware}

After architecture selection, the model is exported as a numerical JavaScript Object Notation (JSON) description. \adaptrevision{The JSON description contains the learned filter coefficients, the polarization-combination rule, and the trigger threshold.
It also specifies the waveform length, sampling rate, and
parameters needed to generate the FPGA implementation.} This representation is the common input to software validation and to the firmware generator (Figure~\ref{fig:firmware_generation_flow}). It contains no training objects and no floating-point covariance data. \auditrevision{The firmware implementation uses the selected whitened-template trigger with a fixed operating threshold of
\(19.56\), calibrated in Section~\ref{subsec:validation_performance}.}

\begin{figure}[tbp]
\centering
\resizebox{0.98\textwidth}{!}{%
\begin{tikzpicture}[
  node distance=8mm and 12mm,
  artifact/.style={draw=scientificblue!65!black, fill=scientificblue!8, align=center, minimum height=8mm, minimum width=28mm},
  process/.style={draw=teal!70!black, fill=teal!10, rounded corners, align=center, minimum height=8mm, minimum width=30mm},
  check/.style={draw=red!70!black, fill=red!9, rounded corners, align=center, minimum height=8mm, minimum width=33mm},
  arr/.style={-{Latex[length=2mm]}, thick}
]
\node[artifact] (model) {selected dual-polarization\\template model};
\node[artifact, right=of model] (jsonnode) {canonical JSON\\model};
\node[process, above right=of jsonnode] (py) {Python\\evaluator};
\node[process, right=of jsonnode] (hls) {JSON-to-HLS\\generator};
\node[artifact, right=of py] (pyout) {reference\\decisions};
\node[artifact, right=of hls] (cpp) {Vitis HLS\\project};
\node[check, right=of cpp] (checks) {HLS/RTL\\agreement checks};
\draw[arr] (model) -- (jsonnode);
\draw[arr] (jsonnode) -- (py);
\draw[arr] (jsonnode) -- (hls);
\draw[arr] (py) -- (pyout);
\draw[arr] (hls) -- (cpp);
\draw[arr] (pyout) -- (checks);
\draw[arr] (cpp) -- (checks);
\end{tikzpicture}
}
\captionsetup{labelfont={color=blue}}
\caption{Firmware-generation workflow. The selected dual-polarization template model is exported to a JSON description that drives both software validation and HLS project generation.}
\captionsetup{labelfont={color=black}}
\label{fig:firmware_generation_flow}
\end{figure}

\subsection{HLS mapping}
\label{sec:json_to_hls}

The HLS block (Figure~\ref{fig:hls_datapath}) scans the full 2016-sample frame and produces a trigger decision. Internally, the design keeps a local sliding response for the two polarization channels, evaluates \adaptrevision{the FIR dot product with unchanged coefficients}, forms absolute responses, computes the local partner maximum within the configured radius, and applies the polarization-sum rule of Equation~\eqref{eq:sum_score}. The maximum frame score is then compared with the calibrated threshold.

The selected model uses one 16-sample template and two channels, corresponding to 32 constant multiplications in the fully parallel local response.  The HLS export uses fixed-point coefficients and fixed accumulator widths configured from the JSON model.  \adaptrevision{The FIR calculation uses unchanged coefficients and contains no floating-point operation or division.}

\begin{figure}[tbp]
\centering
\resizebox{0.98\textwidth}{!}{%
\begin{tikzpicture}[
  node distance=8mm and 10mm,
  data/.style={draw=scientificblue!65!black, fill=scientificblue!8, align=center, minimum height=8mm, minimum width=24mm},
  feature/.style={draw=teal!70!black, fill=teal!10, rounded corners, align=center, minimum height=8mm, minimum width=29mm},
  threshold/.style={draw=red!70!black, fill=red!9, rounded corners, align=center, minimum height=8mm, minimum width=27mm},
  arr/.style={-{Latex[length=2mm]}, thick}
]
\node[data] (frame) {dual-polarization\\ADC frame};
\node[feature, right=of frame] (fir) {two-channel\\16-tap FIR scan};
\node[feature, right=of fir] (pair) {radius-8\\polarization sum};
\node[feature, right=of pair] (max) {frame\\maximum};
\node[threshold, right=of max] (thr) {threshold};
\node[data, right=of thr] (out) {trigger bit};
\draw[arr] (frame) -- (fir);
\draw[arr] (fir) -- (pair);
\draw[arr] (pair) -- (max);
\draw[arr] (max) -- (thr);
\draw[arr] (thr) -- (out);
\end{tikzpicture}
}
\captionsetup{labelfont={color=blue}}
\caption{Generated HLS datapath for the selected model. The implementation scans the synchronized north--south and east--west traces continuously, combines the polarization responses within the coincidence radius, and compares the frame-maximum score with the calibrated threshold.}
\captionsetup{labelfont={color=black}}
\label{fig:hls_datapath}
\end{figure}

The implementation targets an xc7z020-clg400-1 device using Vitis HLS and Vivado 2024.1. The generated design reaches an initiation interval of one clock cycle (\(\mathrm{II}=1\)); after the initial pipeline fill, it accepts one new synchronized dual-polarization sample and produces one new local trigger decision per clock without trigger-induced deadtime. The HLS latency is approximately one frame plus the local pipeline depth; this is expected for the frame-scanning wrapper and is not a throughput limitation. The HLS-generated Verilog was subsequently synthesized, placed, physically optimized, and routed in Vivado with a \SI{5}{\ns} clock constraint and \SI{0.2}{\ns} clock uncertainty.

\begin{table}[tbp]
\centering
\small
\captionsetup{labelfont={color=blue}}
\caption{Final implementation results for the selected dual-polarization
whitened-template trigger on xc7z020-clg400-1.
\auditrevision{Utilization and timing are post-route Vivado results
with the fixed operating threshold of \(19.56\).}}
\captionsetup{labelfont={color=black}}
\label{tab:hls_resources}
\begin{tabularx}{\textwidth}{>{\raggedright\arraybackslash}Xc>{\raggedright\arraybackslash}X}
\toprule
Quantity & Value & Comment \\
\midrule
Target clock & \SI{5.00}{\ns} & \SI{200}{\MHz} target \\
Post-route setup slack & \auditrevision{\(+\SI{0.310}{\ns}\)} & no failing setup endpoints \\
Post-route hold slack & \auditrevision{\(+\SI{0.012}{\ns}\)} & no failing hold endpoints \\
Worst data-path delay & \auditrevision{\SI{4.009}{\ns}} & routed path at \SI{5}{\ns} constraint \\
Latency & \SI{2055}{cycles} & frame scan plus local pipeline \\
Initiation interval & 1 & one synchronized two-polarization sample per clock \\
\SI{18}{\kilo\bit} block RAM & 0 & no table memory required \\
Digital signal-processing (DSP) blocks & 32 & \SI{14.55}{\percent} of target device \\
Slice registers & \auditrevision{5336} & \auditrevision{\SI{5.02}{\percent}} of target device \\
Slice lookup tables & \auditrevision{2379} & \SI{4.47}{\percent} of target device \\
Estimated on-chip power & \auditrevision{\SI{0.302}{\W}} & vectorless estimate \\
\bottomrule
\end{tabularx}
 \footnotesize Power \auditrevision{is a vectorless estimate without a simulation-activity file:} \auditrevision{\SI{0.195}{\W}} \auditrevision{dynamic plus} \SI{0.107}{\W} \auditrevision{static,} \auditrevision{\SI{0.302}{\W}} \auditrevision{total.}
\end{table}

\subsection{RTL validation}
\label{sec:rtl_validation}

\begin{figure}[tbp]
\centering
\begin{subfigure}{0.48\textwidth}
\centering
\includegraphics[width=\textwidth]{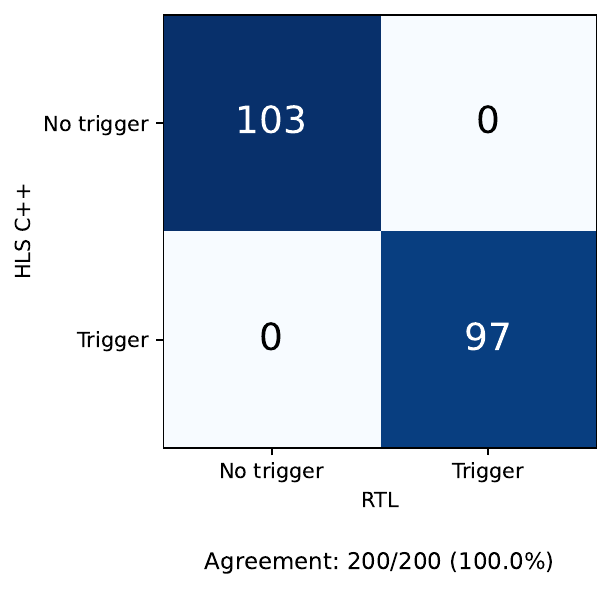}
\captionsetup{labelfont={color=blue}}
\caption{\emph{Validation set 1}.}
\captionsetup{labelfont={color=black}}
\end{subfigure}
\hfill
\begin{subfigure}{0.48\textwidth}
\centering
\includegraphics[width=\textwidth]{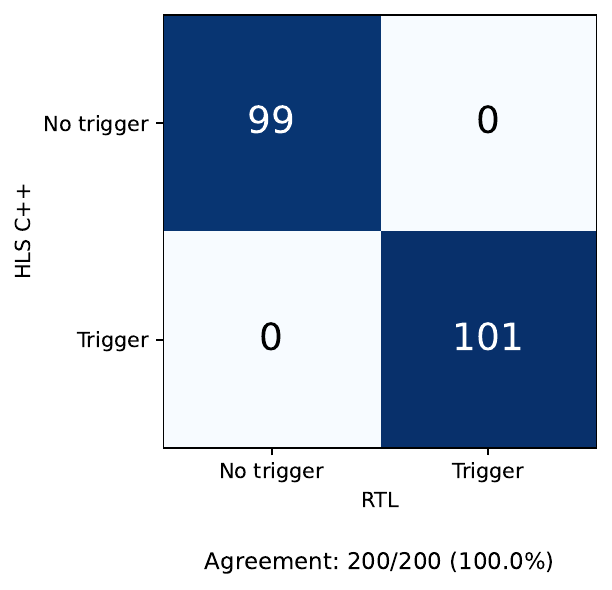}
\caption{\emph{Validation set 2}.}
\end{subfigure}
\caption{HLS and RTL trigger decisions for 200 traces from each validation split. \auditrevision{All 400 RTL decisions agreed with the HLS reference.}}
\label{fig:rtl_confusion}
\end{figure}

The RTL validation compares the trigger bit produced by the generated RTL with the decision from the exported HLS reference.  The test contains 200 paired traces from each validation split: 100 signal traces and 100 background traces.  \adaptrevision{The RTL is tested using \texttt{cocotb}~\cite{HiggsHodgson2013cocotb}, a Python-based hardware verification framework. Its testbench models the one-cycle input-memory timing of the generated RTL and preserves the alignment of the two channels.} After correcting a one-sample channel skew in the initial testbench memory model, the complete 400-trace test was repeated. \auditrevision{All 400 RTL decisions agreed with the HLS reference (Figure~\ref{fig:rtl_confusion}).}

The post-route results use the regenerated final taps and comparator constant. This is an order-of-magnitude implementation estimate rather than a measured power result. The top-level HLS wrapper also has no board-specific pin constraints, which is appropriate for an IP-level implementation but precludes generation of a deployable board bitstream without integration constraints.

\begin{samepage}
\adaptrevision{An additional FPGA prototype integrates the whitened-template trigger with the adaptive-threshold controller described in Section~\ref{subsec:continuous_bolt}, using scores from the preceding approximately \SI{1}{\milli\second} interval to update the operating threshold without changing the FIR coefficients. The integrated design achieves \(\mathrm{II}=1\) and a post-route period of \SI{4.989}{\ns}, satisfying the \SI{200}{\MHz} timing target. Functional validation confirms agreement with the software threshold-setting algorithm.}
\par
\end{samepage}

\section{Discussion and physics interpretation}
\label{sec:discussion}

\subsection{Physics content of the trigger score.}
\label{sec:physics_interpretation}

\begin{figure}[tbp]
\centering
\includegraphics[width=0.98\textwidth]{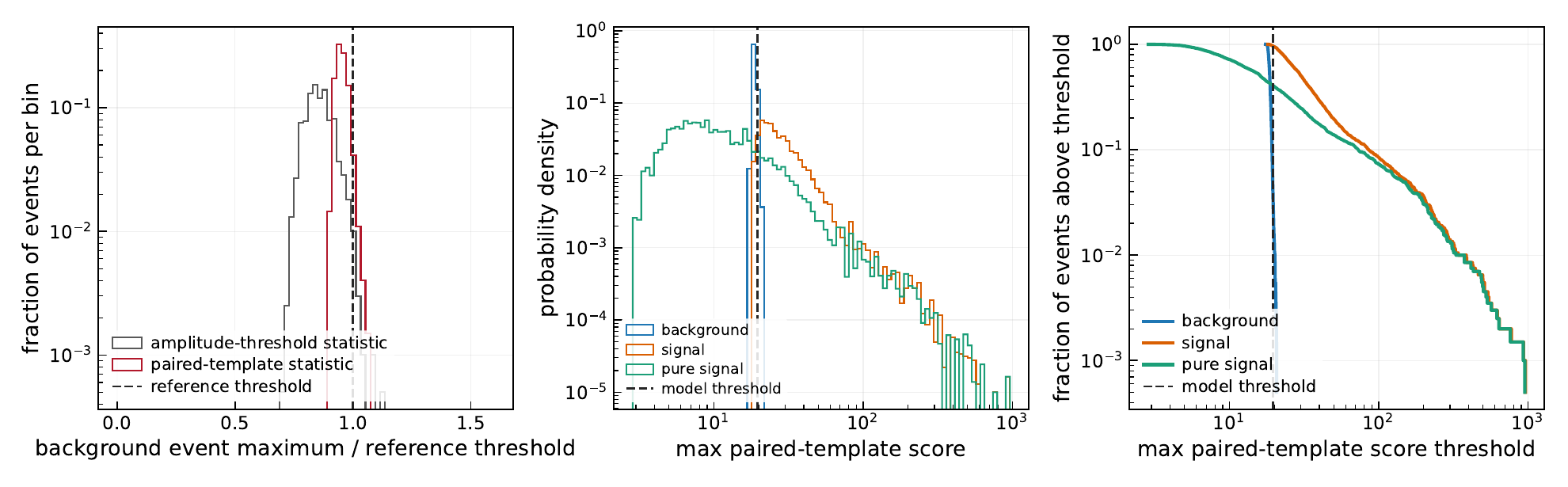}
\captionsetup{labelfont={color=blue}}
\caption{Background trigger-likeness diagnostic on \emph{validation set 2}.  The distributions use the definitions background, signal, and pure signal stated in Section~\ref{sec:data}.  The right panel is a survival curve and is the closest diagnostic to the continuous trigger-rate interpretation.}
\captionsetup{labelfont={color=black}}
\label{fig:background_likeness}
\end{figure}

\begin{figure}[tbp]
\centering
\includegraphics[width=0.90\textwidth]{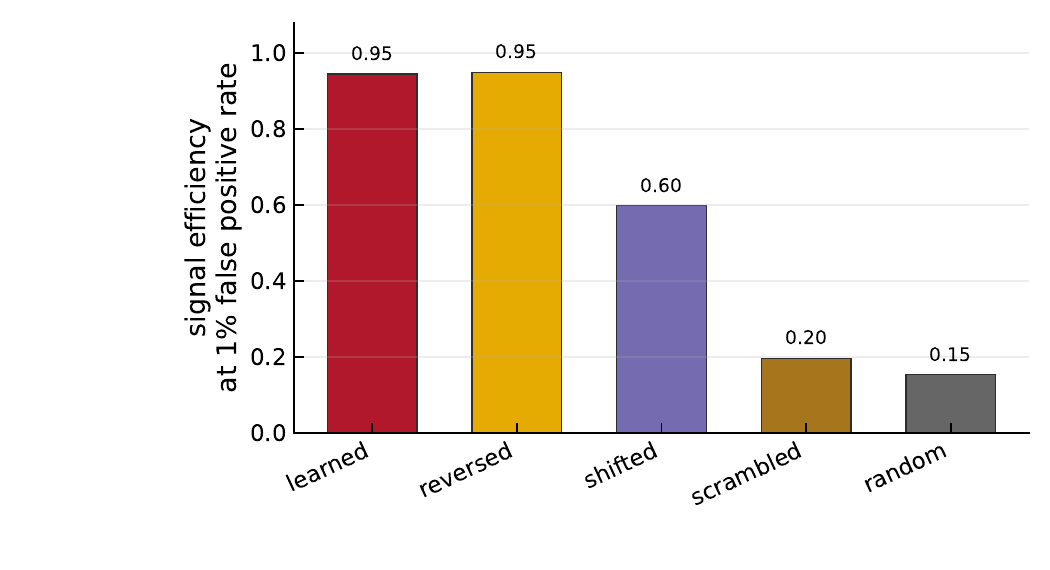}
\captionsetup{labelfont={color=blue}}
\caption{\auditrevision{Pulse-detection efficiency for the learned filter and four altered filters at a \SI{1}{\percent} false-positive rate. Each filter has its own threshold determined from background-only frames. Reversing the coefficients preserves efficiency, while an eight-position circular shift, shuffling, or random coefficients reduce it.}}
\captionsetup{labelfont={color=black}}
\label{fig:template_ablation_operating_point}
\end{figure}

\begin{figure}[tbp]
\centering
\includegraphics[width=0.98\textwidth]{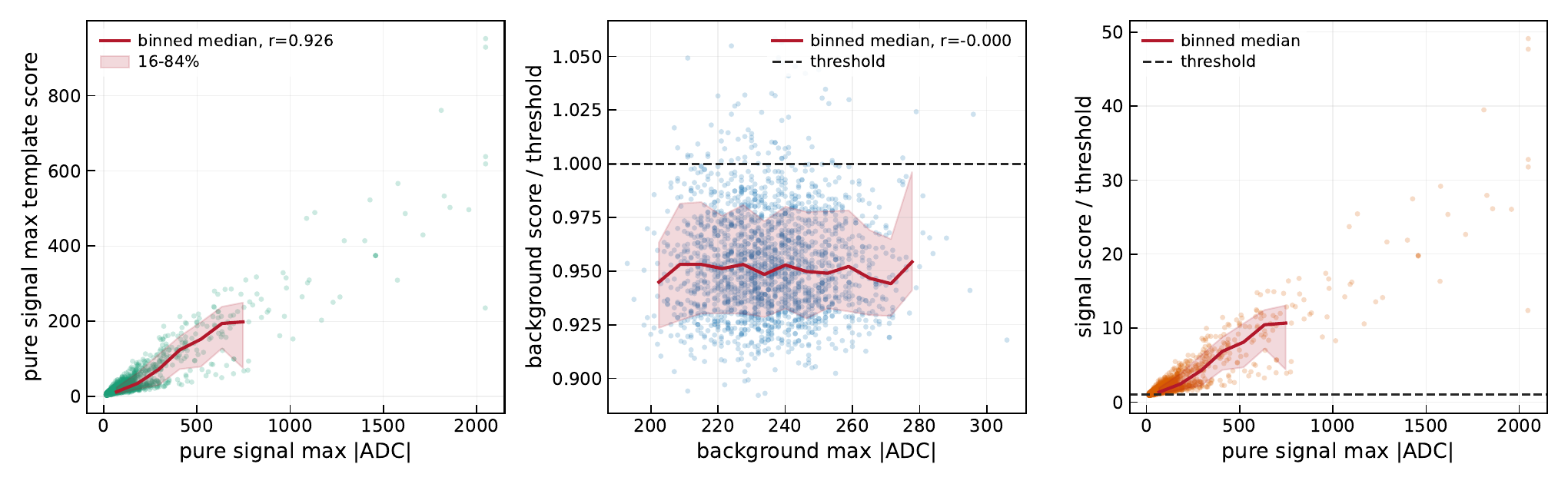}
\captionsetup{labelfont={color=blue}}
\caption{Amplitude-response stability on \emph{validation set 2}.  The pure signal score scales with pure signal amplitude, while the background score margin remains nearly independent of the maximum background ADC amplitude.  The signal score margin stays above threshold for most injected signals.}
\captionsetup{labelfont={color=black}}
\label{fig:amplitude_response_stability}
\end{figure}

\auditrevision{At comparable frame-equivalent candidate rates on \emph{validation set 2}, the selected whitened-template trigger detects 0.96 of injected pulses, compared with 0.40 for the unwhitened template and 0.16 for a peak-amplitude threshold (Table~\ref{tab:operating_points}). Pulse-template matching and background-covariance whitening therefore improve discrimination beyond a cut on the largest ADC sample. Figures~\ref{fig:background_likeness}--\ref{fig:amplitude_response_stability}
examine whether the score exhibits the expected behaviour of a filter sensitive to EAS radio pulses.}

\auditrevision{In Figure~\ref{fig:background_likeness}, the background-only scores cluster near and below the selected threshold, while injected pulses populate much higher scores more often. The distributions still overlap around the threshold, where weak pulses are hardest to distinguish from background. This separation establishes that the score is useful, but does not show whether its performance depends on the learned pulse shape or merely on using a short filter.}

\auditrevision{Figure~\ref{fig:template_ablation_operating_point} addresses that distinction by changing the 16-tap coefficients while retaining the same paired-channel scoring rule. A separate threshold for each filter accepts \SI{1}{\percent} of background-only frames, so differences in signal efficiency cannot be attributed to different background acceptance. This is a diagnostic comparison, separate from the operating threshold in Table~\ref{tab:operating_points}. Reversal tests whether the coefficient sequence works in either temporal orientation. Circularly shifting or shuffling the coefficients retains their values but changes their order, while random coefficients of the same norm provide a short-filter control. The learned and reversed filters each detect about \SI{95}{\percent} of injected pulses. Efficiency falls to about \SI{60}{\percent} after an eight-tap circular shift, \SI{20}{\percent} after shuffling, and \SI{15}{\percent} with random coefficients. The temporal arrangement of the learned taps thus matters substantially, although the near-equal result on reversal does not establish a preferred time direction.}

\auditrevision{Figure~\ref{fig:amplitude_response_stability} separates the response to pulse amplitude from the response to large background samples. For detector-folded pulses before background addition, the largest absolute ADC sample and maximum template score have a Pearson correlation of about 0.93. In background-only frames, the correlation between the largest absolute ADC sample and the maximum score divided by the fixed threshold is near zero, showing little linear association over the sampled amplitude range. These correlations use individual frames; the binned medians in the figure summarize their trends. After pulse injection into measured background, the score relative to threshold generally rises with pure-pulse amplitude, with substantial scatter. The direct peak-amplitude comparison in Table~\ref{tab:operating_points} establishes that this amplitude response does not make the trigger equivalent to an amplitude cut.}

\begin{figure}[tbp]
\centering
\includegraphics[width=0.90\textwidth]{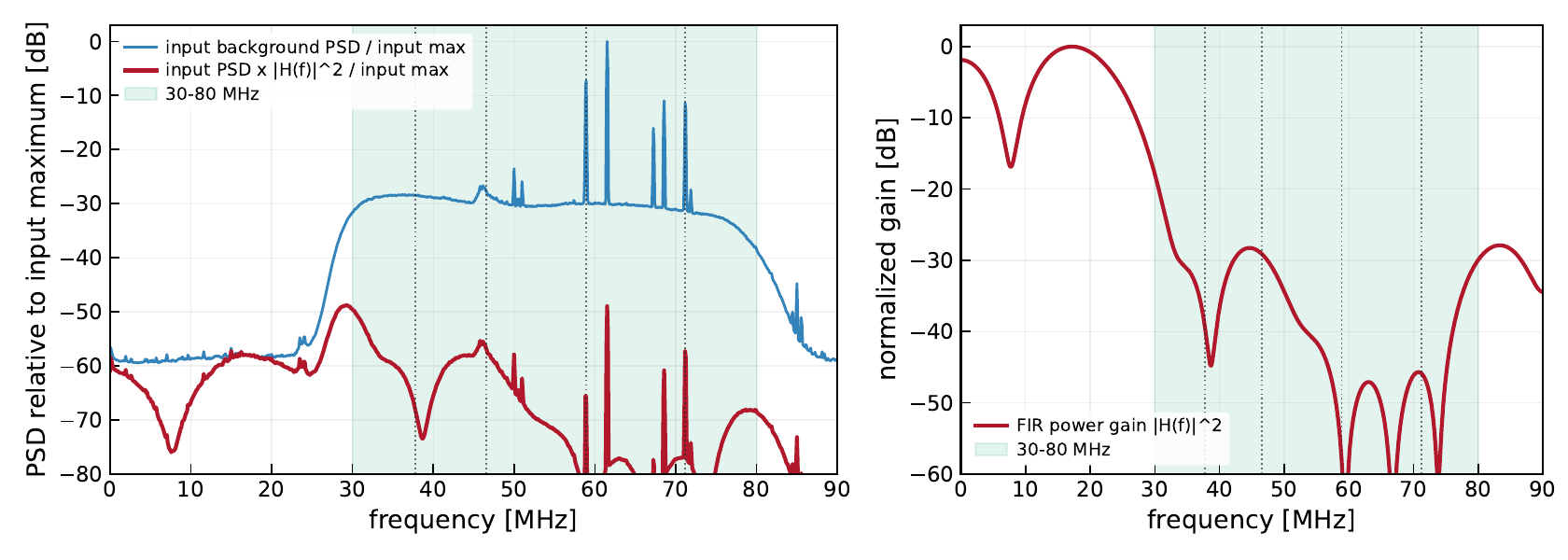}
\captionsetup{labelfont={color=blue}}
\caption{Frequency-domain noise filtering.  The measured input background \auditrevision{power spectral density (PSD)} is compared with the same PSD weighted by the learned FIR power response, \(S_n(f)|H(f)|^2\), using a common normalization to the input PSD maximum.  This avoids independently normalizing the post-filter curve and therefore gives a cleaner view of residual background power.}
\captionsetup{labelfont={color=black}}
\label{fig:frequency_noise_filtering}
\end{figure}

\begin{figure}[tbp]
\centering
\includegraphics[width=0.98\textwidth]{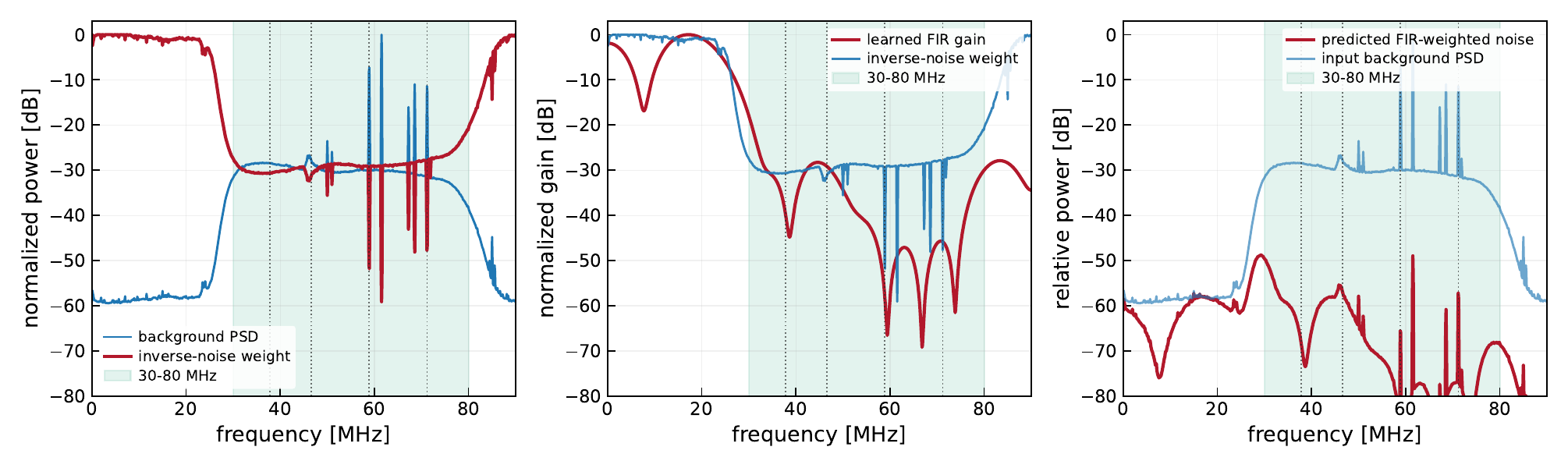}
\captionsetup{labelfont={color=blue}}
\caption{Estimated whitening and residual-noise interpretation.  The inverse-noise weight is estimated from the validation background PSD because the exported model JSON contains the final whitened FIR taps but not the full training covariance matrix.  The figure separates the measured noise spectrum, the inverse-noise weighting, the learned FIR gain, and the predicted FIR-weighted residual background power.}
\captionsetup{labelfont={color=black}}
\label{fig:frequency_whitener_noise_model}
\end{figure}

\auditrevision{The frequency-domain results show how the learned filter treats the measured background. Its input power spectrum has several strong narrow lines, while the FIR gain is low across much of the noisy band and has additional notches near some lines (Figures~\ref{fig:frequency_noise_filtering} and~\ref{fig:frequency_whitener_noise_model}). Multiplying the input spectrum by the FIR power response gives the predicted filtered-background spectrum, \(S_n(f)|H(f)|^2\). On the common scale in Figure~\ref{fig:frequency_noise_filtering}, the lines are reduced and none dominates this prediction. The inverse-noise curve in Figure~\ref{fig:frequency_whitener_noise_model} is estimated from the validation background, whereas the FIR was constructed from the training covariance and the pulse template. Their broadly similar suppression of noisy frequencies is consistent with covariance whitening; the plotted product is a prediction from the measured input spectrum and filter response, not a measured output spectrum.}

\begin{figure}[tbp]
\centering
\includegraphics[width=0.76\textwidth]{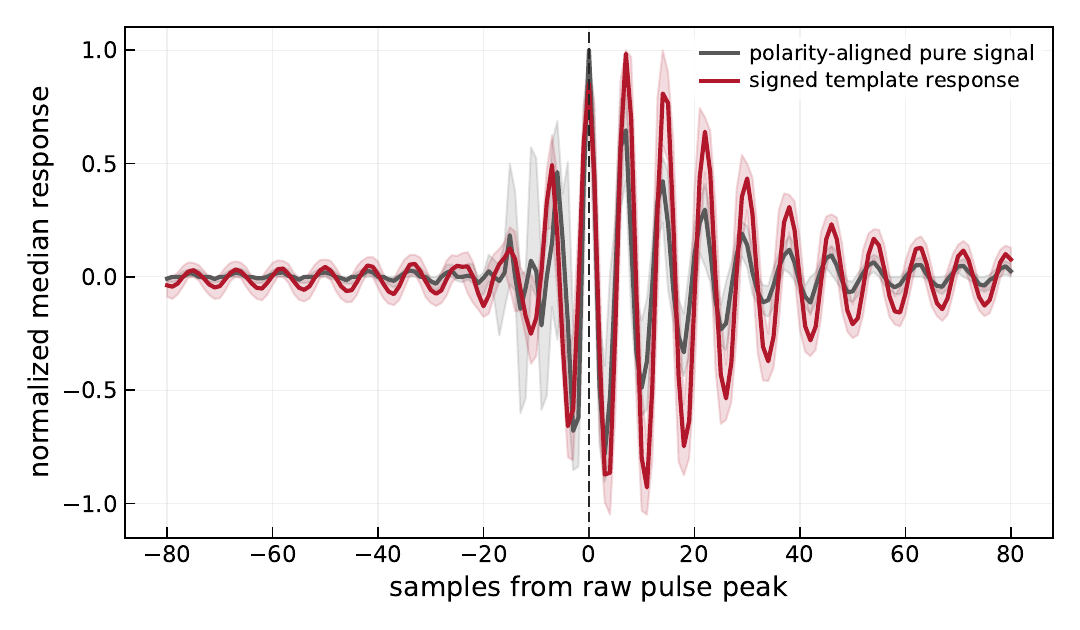}
\captionsetup{labelfont={color=blue}}
\caption{Ringing diagnostic using polarity-aligned pure-signal windows and the signed template response.}
\captionsetup{labelfont={color=black}}
\label{fig:ringing}
\end{figure}

\auditrevision{Figure~\ref{fig:ringing} shows the signed FIR response
around the peak of each pure-signal trace. The median response
exhibits oscillatory lobes that decrease in amplitude away from
the pulse.}

\auditrevision{The comparison with the unwhitened template and
amplitude threshold demonstrates the benefit of incorporating
EAS pulse morphology and measured background correlations.}

\subsection{Scope and limitations}
\label{sec:limitations}

\adaptrevision{The additional BOLT recordings extend the evaluation beyond short background samples from a single reference station by including independent observations from different locations and acquisition periods, together with longer continuous waveforms. They allow the unchanged FIR coefficients to be evaluated under heterogeneous measured interference conditions and preserve the temporal structure needed to assess continuous triggering. This broadens the environmental coverage of the original benchmark, although the finite recording durations and gaps between observations leave long-term variability at each station incompletely characterized.}

\adaptrevision{The method's applicability to another experiment rests on constructing a pulse template and estimating background correlations appropriate to that detector. The particular AERA FIR coefficients incorporate its sampled pulse response and training-background covariance. Different bandwidths, antenna and electronics responses, sampling rates, or radio environments would therefore require corresponding detector-folded signal and background data to construct and validate suitable coefficients. The polarization-combination rule must likewise reflect the available channels and their timing. The compact trigger architecture provides a basis for implementation on other hardware platforms, with resource requirements and timing established for the selected device.}

\adaptrevision{The demonstrated FPGA implementation covers whitened-template filtering, polarization combination, and the trigger decision, and an additional prototype integrates threshold adaptation with unchanged FIR coefficients. Validation covers functional simulation and synthesis/place-and-route, while operation on a physical board remains to be evaluated.}

\section{Summary and outlook}
\label{sec:conclusion}

\adaptrevision{We have demonstrated effective discrimination of air-shower radio pulses against measured backgrounds using a compact, FPGA-compatible whitened-template trigger. Combining EAS pulse morphology with measured background correlations substantially improves weak-pulse detection compared with conventional amplitude thresholding. On the independent benchmark, detector-folded pulses injected into measured background are detected with \SI{96}{\percent} efficiency at a frame-equivalent background trigger rate of \SI{3.12}{\kHz}. At comparable rates, the unwhitened template and amplitude threshold achieve efficiencies of \SI{40}{\percent} and \SI{16}{\percent}, respectively. These comparisons establish the contribution of covariance whitening and pulse-template matching to the observed discrimination.}

\adaptrevision{The advantage for weak pulses persists in later reference-station recordings with the FIR coefficients and threshold unchanged, extending the benchmark result to background acquired at different times. The multi-station BOLT evaluation then examines operation across heterogeneous measured radio environments by transferring the same FIR coefficients and adjusting the threshold to recent background measurements. This evaluation shows how the operating threshold can accommodate changing background conditions and why the temporal clustering of accepted windows must be considered when determining downstream processing capacity.}

\adaptrevision{An FPGA prototype combining the whitened-template trigger with threshold adaptation meets the \SI{200}{\MHz} timing target after place-and-route, demonstrating that the integrated architecture is compatible with continuous processing of the digitized radio signals. The results demonstrate that air-shower pulse morphology and measured background correlations can be exploited for effective radio triggering within the hardware constraints of autonomous stations.}

\FloatBarrier

\section*{Acknowledgments}
The authors gratefully acknowledge the support of the Electronics Laboratory of the Department Physik at Universit\"at Siegen for assistance with the experimental electronics, laboratory infrastructure, and hardware-oriented validation work.  We acknowledge the Pierre Auger Collaboration for providing access to the measured Auger Engineering Radio Array background data and for the simulation and detector-response tools used in the broader radio-trigger development programme.  We particularly thank the Auger Radio Management Team for facilitating access to the radio data used in this study.  We also acknowledge the developers and maintainers of the open-source software used throughout this work, including Python, NumPy, SciPy, scikit-learn, Matplotlib, cocotb, GHDL, Vitis HLS, and Vivado.

\bibliographystyle{JHEP}
\bibliography{sections/09_literature}
\end{document}